\documentclass[aps, prx,superscriptaddress,reprint,showpacs,longbibliography]{revtex4-2}
\usepackage{color}
\usepackage{graphicx,textcomp,amssymb,amsmath,dcolumn}
\usepackage[colorlinks=false, pdfborder={0 0 0}]{hyperref}
\usepackage{siunitx}
\usepackage{comment}
\usepackage{physics}
\usepackage{siunitx}

\usepackage{romannum}

\begin{document}

\title{Spin–valley 0.7 anomaly in bilayer graphene/WSe\textsubscript{2} quantum point contacts} 

\author{Jonas D. Gerber}
\email{gerberjo@phys.ethz.ch}
\affiliation{Solid State Physics Laboratory, ETH Z\"urich, 8093 Z\"urich, Switzerland}
\author{Efe Ersoy}%
\affiliation{Solid State Physics Laboratory, ETH Z\"urich, 8093 Z\"urich, Switzerland}
\author{Michele Masseroni}%
\affiliation{Solid State Physics Laboratory, ETH Z\"urich, 8093 Z\"urich, Switzerland}
\author{Markus Niese}
\affiliation{Solid State Physics Laboratory, ETH Z\"urich, 8093 Z\"urich, Switzerland}
\author{Artem O. Denisov}
 \affiliation{Solid State Physics Laboratory, ETH Z\"urich, 8093 Z\"urich, Switzerland}
\author{Christoph Adam}
     \affiliation{Solid State Physics Laboratory, ETH Z\"urich, 8093 Z\"urich, Switzerland}
\author{Lara Ostertag}
     \affiliation{Solid State Physics Laboratory, ETH Z\"urich, 8093 Z\"urich, Switzerland}
\author{Jessica Richter}
     \affiliation{Solid State Physics Laboratory, ETH Z\"urich, 8093 Z\"urich, Switzerland}
\author{Takashi Taniguchi}
	\affiliation{Research Center for Materials Nanoarchitectonics, National Institute for Materials Science,  1-1 Namiki, Tsukuba 305-0044, Japan}
\author{Kenji Watanabe}
	\affiliation{Research Center for Electronic and Optical Materials, National Institute for Materials Science, 1-1 Namiki, Tsukuba 305-0044, Japan}
\author{Yigal Meir}
\affiliation{Department of Physics, Ben-Gurion University of the Negev, Beer-Sheva 84105, Israel}
\author{Thomas Ihn}
    \affiliation{Solid State Physics Laboratory, ETH Z\"urich, 8093 Z\"urich, Switzerland}
\author{Klaus Ensslin}
    \affiliation{Solid State Physics Laboratory, ETH Z\"urich, 8093 Z\"urich, Switzerland}

\date{\today}

\begin{abstract}

We report a well-resolved 0.7 conductance anomaly at $G=0.7\times(2e^2/h)$ in bilayer graphene/WSe$_2$ quantum point contacts. Proximity-enhanced spin--orbit coupling splits the four-fold ground state of bilayer graphene into well-separated spin-valley locked Kramers doublets. The anomaly emerges between these opposite spin-valley states. Despite fundamentally different band structure and wavefunction characteristics, the temperature and bias phenomenology closely mirror GaAs systems. In contrast, the parallel magnetic field response differs significantly, confirming the central role of valley degrees of freedom. This opens new pathways to study valley-exchange correlation physics in regimes inaccessible to conventional semiconductors.



\end{abstract}
\maketitle

The 0.7 anomaly in transport through 1D quantum point contacts (QPCs) has been observed in many devices made from different materials \cite{Micolich2011}. This feature appears as a conductance shoulder at 0.7 times the height of the first quantized plateau \cite{Thomas1996}. Despite its widespread observation over the past three decades, the origin of the 0.7 anomaly remains under debate, with proposed explanations including spontaneous spin polarization \cite{Thomas1996}, Kondo-like effects \cite{Meir2002, Cronenwett2002}, and interaction-enhanced local spin fluctuations \cite{Bauer2013}. Exploring the anomaly across different material systems can shed light on the underlying mechanism. An interesting direction is to study materials with valley degrees of freedom, where valley exchange interactions may fundamentally alter the correlation physics compared to the purely spin-based mechanisms in conventional semiconductors.

Because III-V semiconductors lack relevant valley degrees of freedom, extensive studies of the 0.7 anomaly in GaAs \cite{Thomas1996,Cronenwett2002, Rokhinson2006} involve purely opposite-spin states. Similar signatures reported in GaN \cite{Chou2005}, InAs \cite{Heedt2016, Lehmann2014}, AlAs \cite{Gunawan2006}, and Ge \cite{Gul2017,Mizokuchi2018, Gao2024} are also attributed to opposite-spin mechanisms. QPCs on Si/SiGe show complex valley-dependent transport with step heights of $4e^2/h$ or $2e^2/h$ depending on confinement \cite{vonPock2016, Scappucci2006, Frucci2010}, where 0.7 anomalies either result from a purely spin-based mechanism for large valley splitting or involve states that cannot be individually distinguished in case of small valley splitting. 

This limitation highlights the need for material platforms that combine valley-involving transport with sufficient energy splitting to resolve individual states. Bilayer graphene (BLG) fulfills these requirements with its spin-valley coupled Kramers pairs protected by time-reversal symmetry. In contrast to same-valley opposite-spin interactions in conventional semiconductors, BLG enables the investigation of opposite-valley opposite-spin mechanisms within a fundamentally different band structure. An open question is whether such distinct microscopic properties alter the phenomenology of correlated transport. However, most BLG QPC studies observe an apparent four-fold degeneracy \cite{Overweg2018PRL, Kraft2018} due to the small intrinsic spin--orbit gap of BLG ($\Delta_\mathrm{SO}\lesssim \SI{80}{\micro eV}$ \cite{Banszerus_SOC_QPC, Kurzmann2021}), leaving spin and valley states energetically unresolved. While some studies have reported intermediate plateaus, their origin remains contested, with competing interpretations attributing them to spin--orbit effects \cite{Banszerus_SOC_QPC} versus interaction-driven mechanisms forming a 0.7 anomaly \cite{Gall2022}. We argue that this ambiguity arises because the intrinsic spin-orbit splitting $\Delta_\mathrm{SO}$ in graphene is small compared to typical QPC longitudinal confinement energies $\hbar\omega_x$ in a saddle-point constriction \cite{Buttiker1990}, preventing clear resolution of individual states.


\begin{figure}[hbt!]
	\includegraphics{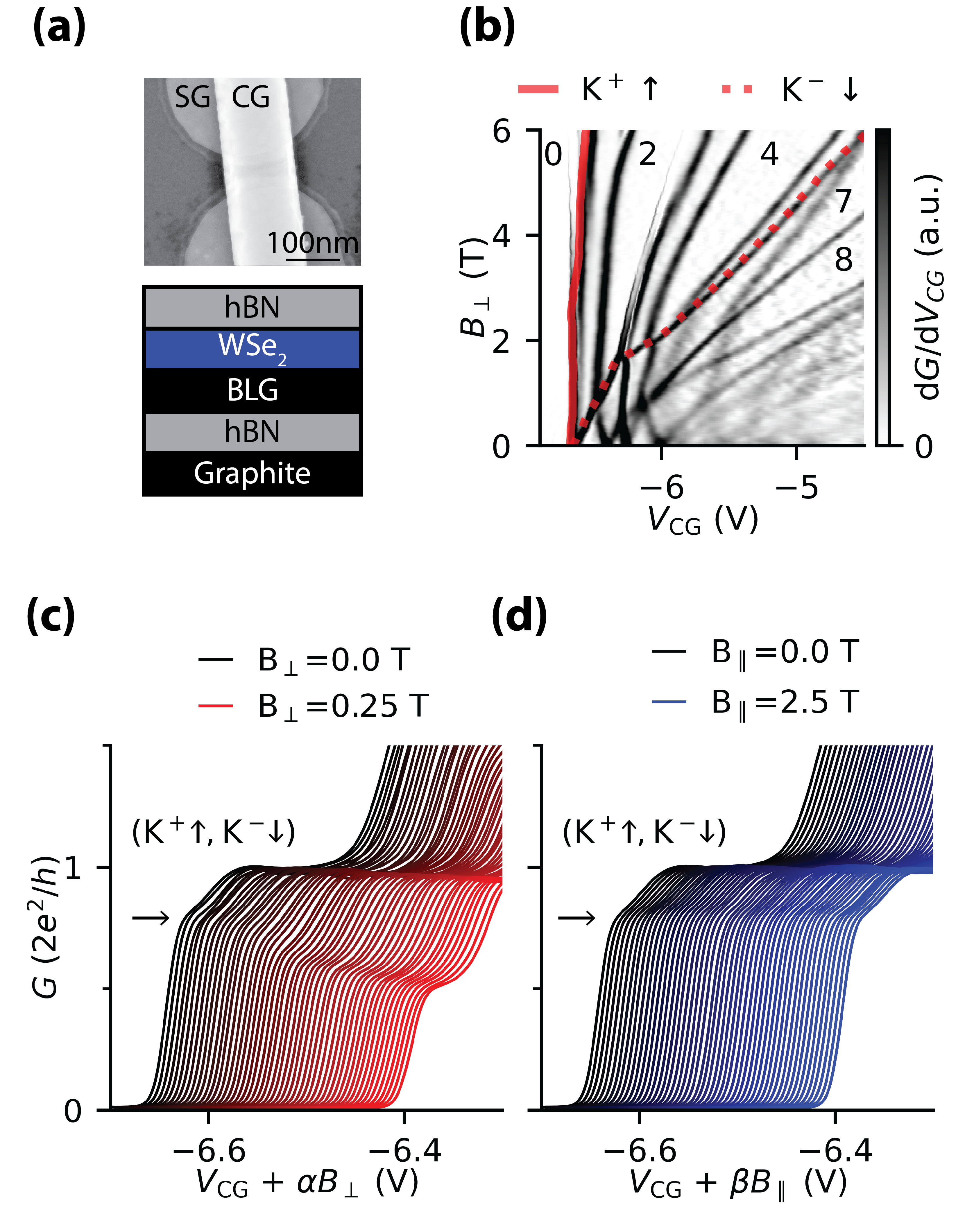}
	\caption{\textbf{(a)} Top-view SEM image and schematic cross-section of the BLG/WSe\textsubscript{2} heterostructure. Split gates (SGs) electrostatically confine the 1D channel, while the channel gate (CG) controls the carrier density.  \textbf{(b)} Transconductance $dG/dV_\mathrm{CG}$ versus $V_\mathrm{CG}$ and $B_\perp$  showing individually resolved transport modes enabled by proximity-enhanced SOC. The legend indicates the evolution of the first subband states $(\mathrm{K}^+\uparrow, \mathrm{K}^-\downarrow)$. \textbf{(c)} Conductance $G$ versus $V_\mathrm{CG}$ for varying perpendicular ($B_\perp$) and \textbf{(d)} parallel magnetic field ($B_\parallel$). Curves are offset horizontally for clarity. The $2e^2/h$ plateau exhibits a pronounced anomaly at roughly $0.7 \times 2e^2/h$ (black arrow) at zero magnetic field. Increasing $B_\perp$ leads to an evolution of the shoulder into a fully quantized $e^2/h$ plateau through valley--Zeeman splitting, while $B_\parallel$ leaves the 0.7 feature unchanged because of spin--valley locking.}
   
	\label{Fig1}
\end{figure}

In this study, we enhance the spin--orbit coupling (SOC) in BLG by interfacing it with WSe\textsubscript{2}, as shown in the schematic heterostructure (Fig.~\ref{Fig1}a). Proximity effects are known to induce significant spin--orbit splitting in BLG/TMD (transition metal dichalcogenide) heterostructures \cite{Gmitra2017, Avsar2014, Wang2015, Yang_2016}. We focus our analysis on a sample with strong proximity-induced spin--orbit coupling ($\Delta_\mathrm{SO}=\SI{1.4}{meV}$ \cite{Gerber2025}), where enhanced energy splitting enables detailed characterization of the 0.7 anomaly. A second sample with weaker SOC ($\Delta_\mathrm{SO}=\SI{0.13}{meV}$) also shows 0.7 signatures, which we analyze in Supplementary Fig.~\ref{FigLowSOC} \cite{supplemental2025}.
Electrostatic gating techniques form the QPC, following techniques established for standard BLG \cite{Overweg2018PRL, Banszerus_SOC_QPC}. Metallic split gates (SGs) confine the n-type 1D channel width, where the channel gate voltage ($V_\mathrm{CG}$) controls the local carrier density.  Additional sample details are provided in the Supplementary Material \cite{supplemental2025}. 

The enhanced $\Delta_\mathrm{SO}$ separates the first degenerate Kramers pair $(\mathrm{K}^+\uparrow, \mathrm{K}^-\downarrow)$ energetically from the second degenerate pair $(\mathrm{K}^-\uparrow, \mathrm{K}^+\downarrow)$. A perpendicular magnetic field $B_\perp$ lifts the remaining degeneracies, enabling single-state resolution as shown in the transconductance plot  (Fig.~\ref{Fig1}b), where black numbers indicate the respective conductance plateaus. Individual mode assignment follows the framework presented in Ref.~\cite{Gerber2025}, where the electronic structure and state ordering in high-SOC BLG/TMD heterostructures have been characterized. The data presented in this manuscript were obtained from the same sample as in Ref.~\cite{Gerber2025}, but in a different parameter regime. 

Figure~\ref{Fig1}c shows the conductance around the first plateau $G_{0} = 2e^{2}/h$ as a function of channel gate voltage for a range of $B_\perp$. This plateau arises from the degenerate ground state Kramers pair $(\mathrm{K}^+\uparrow, \mathrm{K}^-\downarrow)$. Remarkably, we observe a prominent conductance shoulder at $0.7G_0$ (see black arrow in Fig.~\ref{Fig1}c,d) at $B_\perp=0$. Unlike previous reports of intermediate plateaus in BLG, our enhanced energy resolution enables unambiguous identification of the involved states and comprehensive characterization of the anomaly. In a perpendicular magnetic field $B_\perp$, the 0.7 anomaly evolves into a fully resolved $e^2/h$ plateau (see Fig.~\ref{Fig1}c). The extracted g-factor of $\approx 35$ greatly exceeds the pure spin value of 2, proving that the ground state consists of states with opposite valley quantum numbers. Valley g-factors of a similar magnitude have been reported in other BLG QPCs \cite{Lee2020}, far exceeding even interaction-enhanced spin g-factors \cite{Gall2022}.

In contrast to materials like GaAs, increasing the parallel magnetic field $B_\parallel$ does not modify the $0.7G_0$ conductance shoulder in BLG (see Fig.~\ref{Fig1}d). In graphene, at zero magnetic field, the two spin-valley locked states have out-of-plane spins pointing in opposite directions. A parallel field cants both spins simultaneously in plane towards the same direction. As the underlying Hamiltonian is diagonal in the valley subspace, the states remain twofold degenerate. Remarkably, the conductance shoulder remains unchanged even as the spins are no longer strictly antiparallel, indicating the crucial role of valley degrees of freedom in the 0.7 anomaly. The invariance with respect to an application of $B_\parallel$, combined with the valley splitting in $B_\perp$, matches the established behavior for  Kramers pairs in BLG-based heterostructures \cite{Banszerus2021_SpinValley, Kurzmann2021}, confirming the spin-valley ground state $(\mathrm{K}^+\uparrow, \mathrm{K}^-\downarrow)$.


\begin{figure}
	\includegraphics{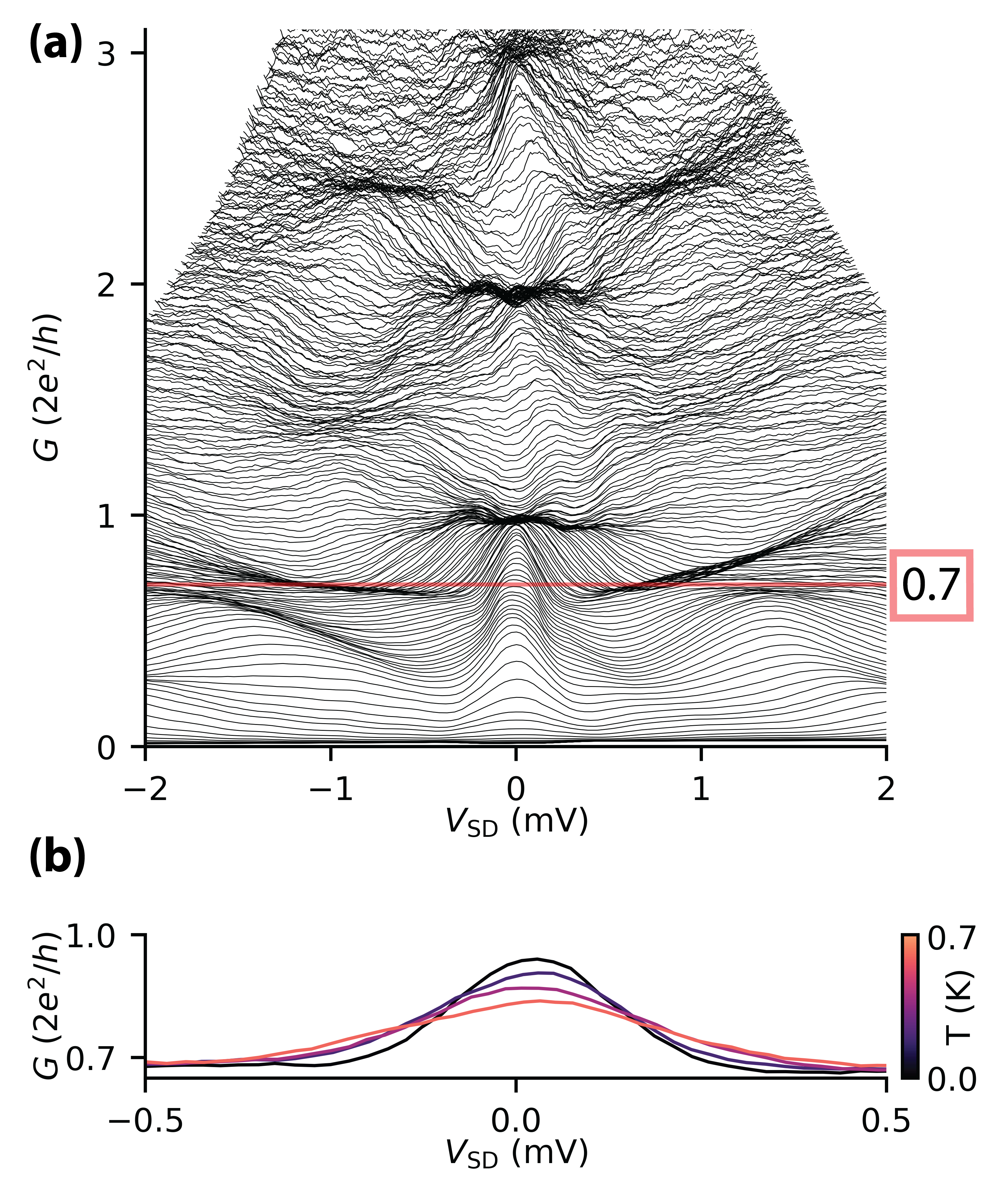}
	\caption{ \textbf{(a)} Differential conductance $G$ versus source-drain bias $V_\mathrm{SD}$ for varying $V_\mathrm{CG}$ with a prominent finite-bias plateau appearing at $0.7 \times 2e^2/h$ and a zero-bias anomaly below the $2e^2/h$ plateau. \textbf{(b)} Temperature dependence of the zero-bias anomaly in conductance $G$ versus $V_\mathrm{SD}$, showing reduced amplitude and increased width at higher temperatures.  }
	\label{Fig2}
\end{figure}

To further establish that this conductance shoulder corresponds to the 0.7 anomaly, we performed detailed finite-bias and temperature characterization, revealing behavior consistent with 0.7 anomaly signatures established in GaAs \cite{Cronenwett2002}.  
Figure \ref{Fig2}a shows the differential conductance $G$ versus source-drain bias $V_\mathrm{SD}$ for varying $V_\mathrm{CG}$, revealing clear $0.7G_0$ plateaus at finite bias and a pronounced zero-bias anomaly below the $2e^2/h$ plateau, both hallmarks of the 0.7 anomaly. The zero-bias anomaly exhibits a characteristic Kondo-like temperature evolution, decreasing in height and increasing in width with increasing temperature (Fig.~\ref{Fig2}b), as observed in GaAs QPCs.

\begin{figure}
	\includegraphics{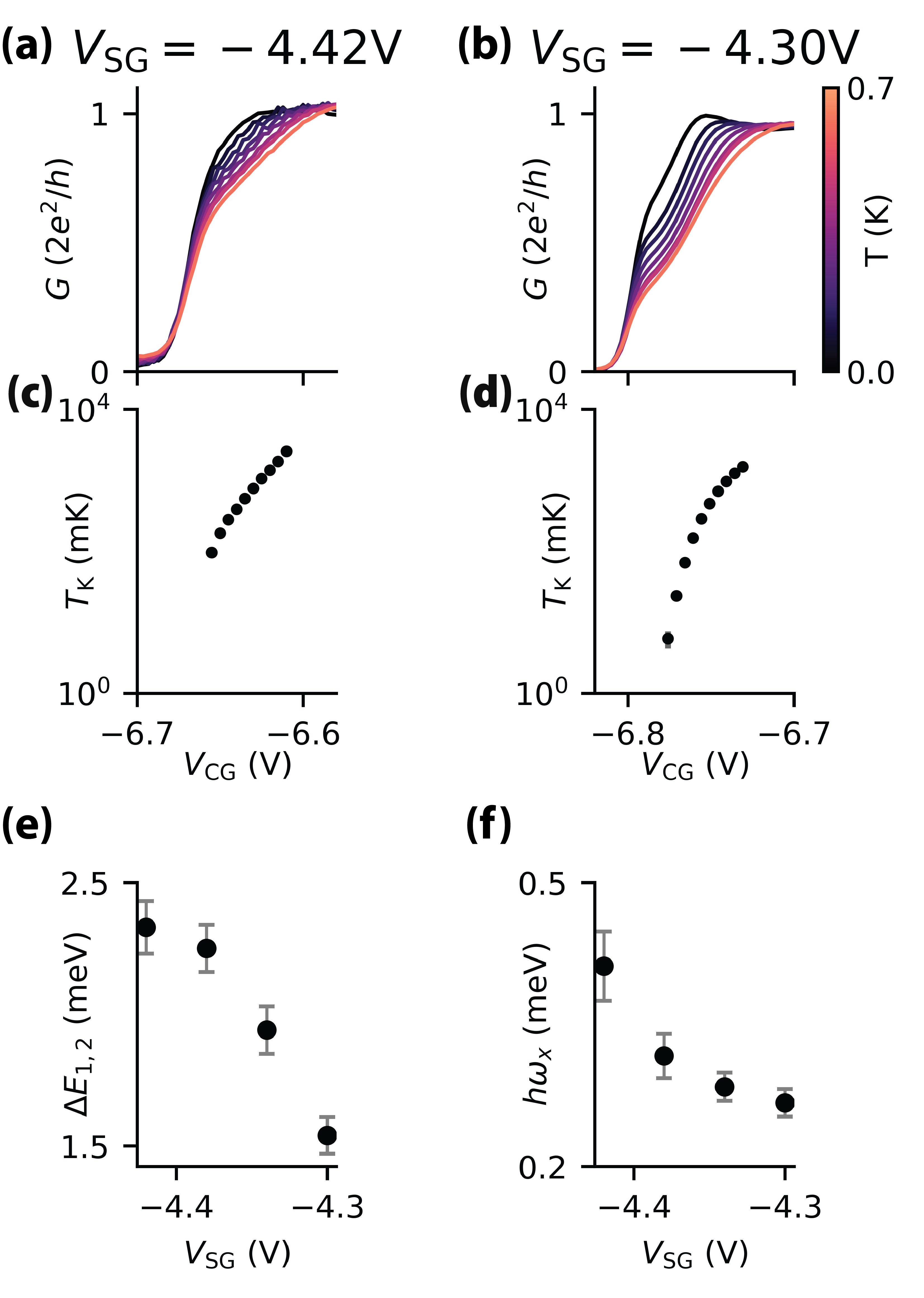}
	\caption{\textbf{(a)}, \textbf{(b)} Temperature-dependent conductance $G$ versus $V_\mathrm{CG}$ for different split gate voltages $V_\mathrm{SG}$. The conductance shows characteristic suppression at elevated temperatures typical of the 0.7 anomaly. \textbf{(c)}, \textbf{(d)} Kondo temperatures $T_K$ extracted from the temperature-dependent fits in (a) and (b) versus $V_\mathrm{CG}$ for both $V_\mathrm{SG}$ values. \textbf{(e)} Subband spacing $\Delta E_{1,2}$ and \textbf{(f)} electrostatic potential curvature $\hbar\omega_{x}$ as functions of $V_\mathrm{SG}$.}
    	\label{Fig3}
\end{figure}

Consequently, the conductance decreases at higher temperatures, as shown for two different split gate voltages $V_\mathrm{SG}$ in Fig.~\ref{Fig3}a,b. With our sample design, we have individual tunability of the channel carrier density via $V_\mathrm{CG}$, and the effective channel width via $V_\mathrm{SG}$. This enables independent tuning of the constriction geometry and 0.7 anomaly properties by varying $V_\mathrm{SG}$. The conductance shoulder becomes more prominent at less negative split gate voltages, where lateral confinement is weaker (Fig.~\ref{Fig3}a,b). We quantify the split gate effect on the 0.7 anomaly by analyzing the temperature dependence of the conductance.  Following established analysis \cite{Cronenwett2002}, we fit $G$ at each $V_\mathrm{CG}$ to the modified Kondo expression $G=G_\mathrm{0}\left[0.5\left(1+(2^{1/s}-1)(T/T_\mathrm{K})^2\right)^{-s}+0.5\right]$ with $G_\mathrm{0}=2e^2/h$ and $s=0.22$. The extracted Kondo temperatures $T_\mathrm{K}$ spanning \SI{10}{mK} to \SI{4}{K} (Fig.~\ref{Fig3}c,d) lie within the GaAs range (0.1~K\,-\,10~K) and exhibit a similar gate voltage dependence \cite{Cronenwett2002}. A complete collection of temperature-dependent conductance traces for various split gate voltages is provided in the Supplementary Material  \cite{supplemental2025} Fig.~\ref{FigSGTemp}.

Electrostatic narrowing of the channel at more negative split gate voltages increases the subband spacing $\Delta E_{1,2}$ as shown in Fig.~\ref{Fig3}e. Analysis using the saddle-point model \cite{Buttiker1990} shows that the split gate voltage also influences the potential curvature in transport direction, characterized by $\hbar\omega_{x}$ (Fig.~\ref{Fig3}f). Smaller $\hbar\omega_{x}$ correlates with a reduced conductance of the shoulder at base temperature and therefore lower minimal Kondo temperatures, which is consistent with results in GaAs \cite{Smith2016}. This systematic control reveals that the spin--valley 0.7 anomaly in BLG is not specific to a particular device geometry. The excellent tunability enables systematic exploration of varying confinements within a single device, eliminating the need for fabricating multiple device geometries. Extraction of $\Delta E_{1,2}$ and $\hbar\omega_{x}$ is detailed in the Supplementary Material \cite{supplemental2025}.

Beyond the standard 0.7 anomaly, we measure similar correlation features at higher subband crossings at finite magnetic fields, including a prominent $1.2G_0$ anomaly (see Supplementary Material \cite{supplemental2025} Fig.~\ref{Fig_0p7anaolg}). These higher-order anomalies, involving spin--valley pairs from different subbands, demonstrate the robustness of interaction effects in our system and mirror the 0.7 analog observed in GaAs \cite{Graham2003}.\\

Although the precise microscopic mechanism of the 0.7 anomaly requires further investigation, our observations clearly demonstrate that the 0.7 anomaly in graphene-based systems forms between spin--valley locked states within a degenerate Kramers pair. This is a fundamental difference from purely spin-based mechanisms in conventional systems, as it involves states with opposite valley quantum numbers. Despite fundamentally different exchange interactions, orbital characteristics, and wavefunctions, the extracted Kondo temperatures, their gate dependence, the dependence of the 0.7 anomaly on the QPC parameters $\Delta E_{1,2}$, $\hbar\omega_x$, and the existence of the 0.7 analog closely match established GaAs systems. The similar energy scales suggest that many-body interactions between opposite spin–valley Kramers pairs in BLG are comparable in strength to purely spin-based interactions in GaAs. This similarity is striking given that short-range valley exchange in BLG produces complex spin-valley multiplet structures in quantum dots \cite{Knothe2020}, yet here it results in a 0.7 phenomenology comparable to conventional semiconductors.

The tunability of BLG systems opens up further prospects for exploring the interplay between bulk and channel correlations. Current theoretical understanding of the 0.7 anomaly assumes unpolarized reservoirs, but BLG's rich phase diagram—including valley/spin polarization and Lifshitz transitions at low densities \cite{Seiler2022, Zhou2022, delaBarrera2022}—could enable probing how bulk phases modify channel correlation physics.   
 Exploring these low-density regimes requires independent displacement field and carrier density control. In our single global back gate design, the large displacement field needed for electrostatic confinement simultaneously drives high carrier densities in the bulk. Independent control would require a dual-gated structure with a global top gate to reduce the carrier density in the bulk while maintaining a high displacement field in the channel for electrostatically defining the QPC.  Investigating the 0.7 anomaly with polarized bulk phases could provide crucial insights, as current theoretical understanding focuses on a polarized channel with unpolarized reservoirs. Such experiments could illuminate the role of bulk-channel coupling in the anomaly's microscopic origin.\\


To summarize, we demonstrated that the 0.7 anomaly in bilayer graphene is based on degenerate spin-valley-locked Kramers pairs. The enhanced SOC separates the two Kramers pairs sufficiently to enable single-state resolution at a finite magnetic field, proving the spin-valley nature of this transport phenomenon. We characterized the temperature, bias, and magnetic field behavior of this anomaly. Crucially, parallel magnetic field measurements reveal the central role of valley degrees of freedom. The anomaly persists even as spins tilt away from antiparallel alignment, demonstrating the essential role of valley exchange interactions. Such valley-dependent mechanisms are absent in conventional semiconductors.  Despite this, the relevant Kondo temperatures, gate voltage dependencies, and the 0.7 analog itself show remarkable similarities to GaAs. This suggests a universality in the many-body correlation physics underlying the 0.7 anomaly: valley-mediated exchange in BLG produces energy scales comparable to spin exchange in GaAs despite fundamentally different orbital characteristics. We further demonstrate excellent in situ tunability of both the subband spacing and the electrostatic potential profile using split and top gates. This, together with its rich phase diagram including valley and spin polarization, establishes bilayer graphene based quantum point contacts as a platform for exploring valley-exchange correlation physics in regimes inaccessible to conventional semiconductors, potentially revealing new insights into the 0.7 anomaly's microscopic origin.\\

\textit{Acknowledgments} - We thank Peter Märki, Thomas Bähler, and the staff of the ETH-cleanroom FIRST for their technical support. We acknowledge support from the European Graphene Flagship Core3 Project, Swiss National Science Foundation via NCCR Quantum Science and Technology, and H2020 European Research Council (ERC) Synergy Grant under Grant Agreement 95154.
K.W. and T.T. acknowledge support from the JSPS KAKENHI (Grant Numbers 21H05233 and 23H02052), the CREST (JPMJCR24A5), JST, and World Premier International Research Center Initiative (WPI), MEXT, Japan. Y. M. acknowledges support from the ISF Breakthrough Research
Grant 737/24. We thank A. Knothe, V. Fal'ko, J. Boddison-Chouinard, L. Gaudreau, and A. F. Young for helpful discussions.

\textit{Data availability} - All data files for the calculations and the figures are available.


\clearpage
\newpage


\begin{thebibliography}{39}%
\makeatletter
\providecommand \@ifxundefined [1]{%
 \@ifx{#1\undefined}
}%
\providecommand \@ifnum [1]{%
 \ifnum #1\expandafter \@firstoftwo
 \else \expandafter \@secondoftwo
 \fi
}%
\providecommand \@ifx [1]{%
 \ifx #1\expandafter \@firstoftwo
 \else \expandafter \@secondoftwo
 \fi
}%
\providecommand \natexlab [1]{#1}%
\providecommand \enquote  [1]{``#1''}%
\providecommand \bibnamefont  [1]{#1}%
\providecommand \bibfnamefont [1]{#1}%
\providecommand \citenamefont [1]{#1}%
\providecommand \href@noop [0]{\@secondoftwo}%
\providecommand \href [0]{\begingroup \@sanitize@url \@href}%
\providecommand \@href[1]{\@@startlink{#1}\@@href}%
\providecommand \@@href[1]{\endgroup#1\@@endlink}%
\providecommand \@sanitize@url [0]{\catcode `\\12\catcode `\$12\catcode `\&12\catcode `\#12\catcode `\^12\catcode `\_12\catcode `\%12\relax}%
\providecommand \@@startlink[1]{}%
\providecommand \@@endlink[0]{}%
\providecommand \url  [0]{\begingroup\@sanitize@url \@url }%
\providecommand \@url [1]{\endgroup\@href {#1}{\urlprefix }}%
\providecommand \urlprefix  [0]{URL }%
\providecommand \Eprint [0]{\href }%
\providecommand \doibase [0]{https://doi.org/}%
\providecommand \selectlanguage [0]{\@gobble}%
\providecommand \bibinfo  [0]{\@secondoftwo}%
\providecommand \bibfield  [0]{\@secondoftwo}%
\providecommand \translation [1]{[#1]}%
\providecommand \BibitemOpen [0]{}%
\providecommand \bibitemStop [0]{}%
\providecommand \bibitemNoStop [0]{.\EOS\space}%
\providecommand \EOS [0]{\spacefactor3000\relax}%
\providecommand \BibitemShut  [1]{\csname bibitem#1\endcsname}%
\let\auto@bib@innerbib\@empty
\bibitem [{\citenamefont {Micolich}(2011)}]{Micolich2011}%
  \BibitemOpen
  \bibfield  {author} {\bibinfo {author} {\bibfnamefont {A.~P.}\ \bibnamefont {Micolich}},\ }\bibfield  {title} {\bibinfo {title} {What lurks below the last plateau: experimental studies of the 0.7 × 2e2/h conductance anomaly in one-dimensional systems},\ }\href {https://doi.org/10.1088/0953-8984/23/44/443201} {\bibfield  {journal} {\bibinfo  {journal} {Journal of Physics: Condensed Matter}\ }\textbf {\bibinfo {volume} {23}},\ \bibinfo {pages} {443201} (\bibinfo {year} {2011})}\BibitemShut {NoStop}%
\bibitem [{\citenamefont {Thomas}\ \emph {et~al.}(1996)\citenamefont {Thomas}, \citenamefont {Nicholls}, \citenamefont {Simmons}, \citenamefont {Pepper}, \citenamefont {Mace},\ and\ \citenamefont {Ritchie}}]{Thomas1996}%
  \BibitemOpen
  \bibfield  {author} {\bibinfo {author} {\bibfnamefont {K.~J.}\ \bibnamefont {Thomas}}, \bibinfo {author} {\bibfnamefont {J.~T.}\ \bibnamefont {Nicholls}}, \bibinfo {author} {\bibfnamefont {M.~Y.}\ \bibnamefont {Simmons}}, \bibinfo {author} {\bibfnamefont {M.}~\bibnamefont {Pepper}}, \bibinfo {author} {\bibfnamefont {D.~R.}\ \bibnamefont {Mace}},\ and\ \bibinfo {author} {\bibfnamefont {D.~A.}\ \bibnamefont {Ritchie}},\ }\bibfield  {title} {\bibinfo {title} {Possible spin polarization in a one-dimensional electron gas},\ }\href {https://doi.org/10.1103/PhysRevLett.77.135} {\bibfield  {journal} {\bibinfo  {journal} {Phys. Rev. Lett.}\ }\textbf {\bibinfo {volume} {77}},\ \bibinfo {pages} {135} (\bibinfo {year} {1996})}\BibitemShut {NoStop}%
\bibitem [{\citenamefont {Meir}\ \emph {et~al.}(2002)\citenamefont {Meir}, \citenamefont {Hirose},\ and\ \citenamefont {Wingreen}}]{Meir2002}%
  \BibitemOpen
  \bibfield  {author} {\bibinfo {author} {\bibfnamefont {Y.}~\bibnamefont {Meir}}, \bibinfo {author} {\bibfnamefont {K.}~\bibnamefont {Hirose}},\ and\ \bibinfo {author} {\bibfnamefont {N.~S.}\ \bibnamefont {Wingreen}},\ }\bibfield  {title} {\bibinfo {title} {Kondo model for the ``0.7 anomaly'' in transport through a quantum point contact},\ }\href {https://doi.org/10.1103/PhysRevLett.89.196802} {\bibfield  {journal} {\bibinfo  {journal} {Phys. Rev. Lett.}\ }\textbf {\bibinfo {volume} {89}},\ \bibinfo {pages} {196802} (\bibinfo {year} {2002})}\BibitemShut {NoStop}%
\bibitem [{\citenamefont {Cronenwett}\ \emph {et~al.}(2002)\citenamefont {Cronenwett}, \citenamefont {Lynch}, \citenamefont {Goldhaber-Gordon}, \citenamefont {Kouwenhoven}, \citenamefont {Marcus}, \citenamefont {Hirose}, \citenamefont {Wingreen},\ and\ \citenamefont {Umansky}}]{Cronenwett2002}%
  \BibitemOpen
  \bibfield  {author} {\bibinfo {author} {\bibfnamefont {S.~M.}\ \bibnamefont {Cronenwett}}, \bibinfo {author} {\bibfnamefont {H.~J.}\ \bibnamefont {Lynch}}, \bibinfo {author} {\bibfnamefont {D.}~\bibnamefont {Goldhaber-Gordon}}, \bibinfo {author} {\bibfnamefont {L.~P.}\ \bibnamefont {Kouwenhoven}}, \bibinfo {author} {\bibfnamefont {C.~M.}\ \bibnamefont {Marcus}}, \bibinfo {author} {\bibfnamefont {K.}~\bibnamefont {Hirose}}, \bibinfo {author} {\bibfnamefont {N.~S.}\ \bibnamefont {Wingreen}},\ and\ \bibinfo {author} {\bibfnamefont {V.}~\bibnamefont {Umansky}},\ }\bibfield  {title} {\bibinfo {title} {Low-temperature fate of the 0.7 structure in a point contact: A kondo-like correlated state in an open system},\ }\href {https://doi.org/10.1103/PhysRevLett.88.226805} {\bibfield  {journal} {\bibinfo  {journal} {Phys. Rev. Lett.}\ }\textbf {\bibinfo {volume} {88}},\ \bibinfo {pages} {226805} (\bibinfo {year} {2002})}\BibitemShut {NoStop}%
\bibitem [{\citenamefont {Bauer}\ \emph {et~al.}(2013)\citenamefont {Bauer}, \citenamefont {Heyder}, \citenamefont {Schubert}, \citenamefont {Borowsky}, \citenamefont {Taubert}, \citenamefont {Bruognolo}, \citenamefont {Schuh}, \citenamefont {Wegscheider}, \citenamefont {von Delft},\ and\ \citenamefont {Ludwig}}]{Bauer2013}%
  \BibitemOpen
  \bibfield  {author} {\bibinfo {author} {\bibfnamefont {F.}~\bibnamefont {Bauer}}, \bibinfo {author} {\bibfnamefont {J.}~\bibnamefont {Heyder}}, \bibinfo {author} {\bibfnamefont {E.}~\bibnamefont {Schubert}}, \bibinfo {author} {\bibfnamefont {D.}~\bibnamefont {Borowsky}}, \bibinfo {author} {\bibfnamefont {D.}~\bibnamefont {Taubert}}, \bibinfo {author} {\bibfnamefont {B.}~\bibnamefont {Bruognolo}}, \bibinfo {author} {\bibfnamefont {D.}~\bibnamefont {Schuh}}, \bibinfo {author} {\bibfnamefont {W.}~\bibnamefont {Wegscheider}}, \bibinfo {author} {\bibfnamefont {J.}~\bibnamefont {von Delft}},\ and\ \bibinfo {author} {\bibfnamefont {S.}~\bibnamefont {Ludwig}},\ }\bibfield  {title} {\bibinfo {title} {Microscopic origin of the `0.7-anomaly' in quantum point contacts},\ }\href {https://doi.org/10.1038/nature12421} {\bibfield  {journal} {\bibinfo  {journal} {Nature}\ }\textbf {\bibinfo {volume} {501}},\ \bibinfo {pages} {73} (\bibinfo {year} {2013})}\BibitemShut {NoStop}%
\bibitem [{\citenamefont {Rokhinson}\ \emph {et~al.}(2006)\citenamefont {Rokhinson}, \citenamefont {Pfeiffer},\ and\ \citenamefont {West}}]{Rokhinson2006}%
  \BibitemOpen
  \bibfield  {author} {\bibinfo {author} {\bibfnamefont {L.~P.}\ \bibnamefont {Rokhinson}}, \bibinfo {author} {\bibfnamefont {L.~N.}\ \bibnamefont {Pfeiffer}},\ and\ \bibinfo {author} {\bibfnamefont {K.~W.}\ \bibnamefont {West}},\ }\bibfield  {title} {\bibinfo {title} {Spontaneous spin polarization in quantum point contacts},\ }\href {https://doi.org/10.1103/PhysRevLett.96.156602} {\bibfield  {journal} {\bibinfo  {journal} {Phys. Rev. Lett.}\ }\textbf {\bibinfo {volume} {96}},\ \bibinfo {pages} {156602} (\bibinfo {year} {2006})}\BibitemShut {NoStop}%
\bibitem [{\citenamefont {Chou}\ \emph {et~al.}(2005)\citenamefont {Chou}, \citenamefont {Lüscher}, \citenamefont {Goldhaber-Gordon}, \citenamefont {Manfra}, \citenamefont {Sergent}, \citenamefont {West},\ and\ \citenamefont {Molnar}}]{Chou2005}%
  \BibitemOpen
  \bibfield  {author} {\bibinfo {author} {\bibfnamefont {H.~T.}\ \bibnamefont {Chou}}, \bibinfo {author} {\bibfnamefont {S.}~\bibnamefont {Lüscher}}, \bibinfo {author} {\bibfnamefont {D.}~\bibnamefont {Goldhaber-Gordon}}, \bibinfo {author} {\bibfnamefont {M.~J.}\ \bibnamefont {Manfra}}, \bibinfo {author} {\bibfnamefont {A.~M.}\ \bibnamefont {Sergent}}, \bibinfo {author} {\bibfnamefont {K.~W.}\ \bibnamefont {West}},\ and\ \bibinfo {author} {\bibfnamefont {R.~J.}\ \bibnamefont {Molnar}},\ }\bibfield  {title} {\bibinfo {title} {High-quality quantum point contacts in {GaN/AlGaN} heterostructures},\ }\href {https://doi.org/10.1063/1.1862339} {\bibfield  {journal} {\bibinfo  {journal} {Applied Physics Letters}\ }\textbf {\bibinfo {volume} {86}},\ \bibinfo {pages} {073108} (\bibinfo {year} {2005})}\BibitemShut {NoStop}%
\bibitem [{\citenamefont {Heedt}\ \emph {et~al.}(2016)\citenamefont {Heedt}, \citenamefont {Prost}, \citenamefont {Schubert}, \citenamefont {Gr{\"u}tzmacher},\ and\ \citenamefont {Sch{\"a}pers}}]{Heedt2016}%
  \BibitemOpen
  \bibfield  {author} {\bibinfo {author} {\bibfnamefont {S.}~\bibnamefont {Heedt}}, \bibinfo {author} {\bibfnamefont {W.}~\bibnamefont {Prost}}, \bibinfo {author} {\bibfnamefont {J.}~\bibnamefont {Schubert}}, \bibinfo {author} {\bibfnamefont {D.}~\bibnamefont {Gr{\"u}tzmacher}},\ and\ \bibinfo {author} {\bibfnamefont {T.}~\bibnamefont {Sch{\"a}pers}},\ }\bibfield  {title} {\bibinfo {title} {Ballistic transport and exchange interaction in inas nanowire quantum point contacts},\ }\href {https://doi.org/10.1021/acs.nanolett.6b00414} {\bibfield  {journal} {\bibinfo  {journal} {Nano Letters}\ }\textbf {\bibinfo {volume} {16}},\ \bibinfo {pages} {3116} (\bibinfo {year} {2016})}\BibitemShut {NoStop}%
\bibitem [{\citenamefont {Lehmann}\ \emph {et~al.}(2014)\citenamefont {Lehmann}, \citenamefont {Benter}, \citenamefont {von Ahnen}, \citenamefont {Jacob}, \citenamefont {Matsuyama}, \citenamefont {Merkt}, \citenamefont {Kunze}, \citenamefont {Wieck}, \citenamefont {Reuter}, \citenamefont {Heyn},\ and\ \citenamefont {Hansen}}]{Lehmann2014}%
  \BibitemOpen
  \bibfield  {author} {\bibinfo {author} {\bibfnamefont {H.}~\bibnamefont {Lehmann}}, \bibinfo {author} {\bibfnamefont {T.}~\bibnamefont {Benter}}, \bibinfo {author} {\bibfnamefont {I.}~\bibnamefont {von Ahnen}}, \bibinfo {author} {\bibfnamefont {J.}~\bibnamefont {Jacob}}, \bibinfo {author} {\bibfnamefont {T.}~\bibnamefont {Matsuyama}}, \bibinfo {author} {\bibfnamefont {U.}~\bibnamefont {Merkt}}, \bibinfo {author} {\bibfnamefont {U.}~\bibnamefont {Kunze}}, \bibinfo {author} {\bibfnamefont {A.~D.}\ \bibnamefont {Wieck}}, \bibinfo {author} {\bibfnamefont {D.}~\bibnamefont {Reuter}}, \bibinfo {author} {\bibfnamefont {C.}~\bibnamefont {Heyn}},\ and\ \bibinfo {author} {\bibfnamefont {W.}~\bibnamefont {Hansen}},\ }\bibfield  {title} {\bibinfo {title} {Spin-resolved conductance quantization in inas},\ }\href {https://doi.org/10.1088/0268-1242/29/7/075010} {\bibfield  {journal} {\bibinfo  {journal} {Semiconductor Science and Technology}\ }\textbf {\bibinfo {volume} {29}},\ \bibinfo {pages} {075010} (\bibinfo {year}
  {2014})}\BibitemShut {NoStop}%
\bibitem [{\citenamefont {Gunawan}\ \emph {et~al.}(2006)\citenamefont {Gunawan}, \citenamefont {Habib}, \citenamefont {De~Poortere},\ and\ \citenamefont {Shayegan}}]{Gunawan2006}%
  \BibitemOpen
  \bibfield  {author} {\bibinfo {author} {\bibfnamefont {O.}~\bibnamefont {Gunawan}}, \bibinfo {author} {\bibfnamefont {B.}~\bibnamefont {Habib}}, \bibinfo {author} {\bibfnamefont {E.~P.}\ \bibnamefont {De~Poortere}},\ and\ \bibinfo {author} {\bibfnamefont {M.}~\bibnamefont {Shayegan}},\ }\bibfield  {title} {\bibinfo {title} {Quantized conductance in an alas two-dimensional electron system quantum point contact},\ }\href {https://doi.org/10.1103/PhysRevB.74.155436} {\bibfield  {journal} {\bibinfo  {journal} {Phys. Rev. B}\ }\textbf {\bibinfo {volume} {74}},\ \bibinfo {pages} {155436} (\bibinfo {year} {2006})}\BibitemShut {NoStop}%
\bibitem [{\citenamefont {Gul}\ \emph {et~al.}(2017)\citenamefont {Gul}, \citenamefont {Holmes}, \citenamefont {Newton}, \citenamefont {Ellis}, \citenamefont {Morrison}, \citenamefont {Pepper}, \citenamefont {Barnes},\ and\ \citenamefont {Myronov}}]{Gul2017}%
  \BibitemOpen
  \bibfield  {author} {\bibinfo {author} {\bibfnamefont {Y.}~\bibnamefont {Gul}}, \bibinfo {author} {\bibfnamefont {S.~N.}\ \bibnamefont {Holmes}}, \bibinfo {author} {\bibfnamefont {P.~J.}\ \bibnamefont {Newton}}, \bibinfo {author} {\bibfnamefont {D.~J.~P.}\ \bibnamefont {Ellis}}, \bibinfo {author} {\bibfnamefont {C.}~\bibnamefont {Morrison}}, \bibinfo {author} {\bibfnamefont {M.}~\bibnamefont {Pepper}}, \bibinfo {author} {\bibfnamefont {C.~H.~W.}\ \bibnamefont {Barnes}},\ and\ \bibinfo {author} {\bibfnamefont {M.}~\bibnamefont {Myronov}},\ }\bibfield  {title} {\bibinfo {title} {Quantum ballistic transport in strained epitaxial germanium},\ }\href {https://doi.org/10.1063/1.5008969} {\bibfield  {journal} {\bibinfo  {journal} {Applied Physics Letters}\ }\textbf {\bibinfo {volume} {111}},\ \bibinfo {pages} {233512} (\bibinfo {year} {2017})}\BibitemShut {NoStop}%
\bibitem [{\citenamefont {Mizokuchi}\ \emph {et~al.}(2018)\citenamefont {Mizokuchi}, \citenamefont {Maurand}, \citenamefont {Vigneau}, \citenamefont {Myronov},\ and\ \citenamefont {De~Franceschi}}]{Mizokuchi2018}%
  \BibitemOpen
  \bibfield  {author} {\bibinfo {author} {\bibfnamefont {R.}~\bibnamefont {Mizokuchi}}, \bibinfo {author} {\bibfnamefont {R.}~\bibnamefont {Maurand}}, \bibinfo {author} {\bibfnamefont {F.}~\bibnamefont {Vigneau}}, \bibinfo {author} {\bibfnamefont {M.}~\bibnamefont {Myronov}},\ and\ \bibinfo {author} {\bibfnamefont {S.}~\bibnamefont {De~Franceschi}},\ }\bibfield  {title} {\bibinfo {title} {Ballistic one-dimensional holes with strong g-factor anisotropy in germanium},\ }\href {https://doi.org/10.1021/acs.nanolett.8b01457} {\bibfield  {journal} {\bibinfo  {journal} {Nano Letters}\ }\textbf {\bibinfo {volume} {18}},\ \bibinfo {pages} {4861} (\bibinfo {year} {2018})}\BibitemShut {NoStop}%
\bibitem [{\citenamefont {Gao}\ \emph {et~al.}(2024)\citenamefont {Gao}, \citenamefont {Kong}, \citenamefont {Zhang}, \citenamefont {Luo}, \citenamefont {Su}, \citenamefont {Liu}, \citenamefont {Wang}, \citenamefont {Wang},\ and\ \citenamefont {Xu}}]{Gao2024}%
  \BibitemOpen
  \bibfield  {author} {\bibinfo {author} {\bibfnamefont {H.}~\bibnamefont {Gao}}, \bibinfo {author} {\bibfnamefont {Z.-Z.}\ \bibnamefont {Kong}}, \bibinfo {author} {\bibfnamefont {P.}~\bibnamefont {Zhang}}, \bibinfo {author} {\bibfnamefont {Y.}~\bibnamefont {Luo}}, \bibinfo {author} {\bibfnamefont {H.}~\bibnamefont {Su}}, \bibinfo {author} {\bibfnamefont {X.-F.}\ \bibnamefont {Liu}}, \bibinfo {author} {\bibfnamefont {G.-L.}\ \bibnamefont {Wang}}, \bibinfo {author} {\bibfnamefont {J.-Y.}\ \bibnamefont {Wang}},\ and\ \bibinfo {author} {\bibfnamefont {H.~Q.}\ \bibnamefont {Xu}},\ }\bibfield  {title} {\bibinfo {title} {Gate-defined quantum point contacts in a germanium quantum well},\ }\href {https://doi.org/10.1039/D4NR00712C} {\bibfield  {journal} {\bibinfo  {journal} {Nanoscale}\ }\textbf {\bibinfo {volume} {16}},\ \bibinfo {pages} {10333} (\bibinfo {year} {2024})}\BibitemShut {NoStop}%
\bibitem [{\citenamefont {von Pock}\ \emph {et~al.}(2016)\citenamefont {von Pock}, \citenamefont {Salloch}, \citenamefont {Qiao}, \citenamefont {Wieser}, \citenamefont {Hackbarth},\ and\ \citenamefont {Kunze}}]{vonPock2016}%
  \BibitemOpen
  \bibfield  {author} {\bibinfo {author} {\bibfnamefont {J.~F.}\ \bibnamefont {von Pock}}, \bibinfo {author} {\bibfnamefont {D.}~\bibnamefont {Salloch}}, \bibinfo {author} {\bibfnamefont {G.}~\bibnamefont {Qiao}}, \bibinfo {author} {\bibfnamefont {U.}~\bibnamefont {Wieser}}, \bibinfo {author} {\bibfnamefont {T.}~\bibnamefont {Hackbarth}},\ and\ \bibinfo {author} {\bibfnamefont {U.}~\bibnamefont {Kunze}},\ }\bibfield  {title} {\bibinfo {title} {Quantization and anomalous structures in the conductance of {Si/SiGe} quantum point contacts},\ }\href {https://doi.org/10.1063/1.4945116} {\bibfield  {journal} {\bibinfo  {journal} {Journal of Applied Physics}\ }\textbf {\bibinfo {volume} {119}},\ \bibinfo {pages} {134306} (\bibinfo {year} {2016})}\BibitemShut {NoStop}%
\bibitem [{\citenamefont {Scappucci}\ \emph {et~al.}(2006)\citenamefont {Scappucci}, \citenamefont {Gaspare}, \citenamefont {Giovine}, \citenamefont {Notargiacomo}, \citenamefont {Leoni},\ and\ \citenamefont {Evangelisti}}]{Scappucci2006}%
  \BibitemOpen
  \bibfield  {author} {\bibinfo {author} {\bibfnamefont {G.}~\bibnamefont {Scappucci}}, \bibinfo {author} {\bibfnamefont {L.~D.}\ \bibnamefont {Gaspare}}, \bibinfo {author} {\bibfnamefont {E.}~\bibnamefont {Giovine}}, \bibinfo {author} {\bibfnamefont {A.}~\bibnamefont {Notargiacomo}}, \bibinfo {author} {\bibfnamefont {R.}~\bibnamefont {Leoni}},\ and\ \bibinfo {author} {\bibfnamefont {F.}~\bibnamefont {Evangelisti}},\ }\bibfield  {title} {\bibinfo {title} {Conductance quantization in etched {Si/SiGe} quantum point contacts},\ }\href {https://doi.org/10.1103/PhysRevB.74.035321} {\bibfield  {journal} {\bibinfo  {journal} {Phys. Rev. B}\ }\textbf {\bibinfo {volume} {74}},\ \bibinfo {pages} {035321} (\bibinfo {year} {2006})}\BibitemShut {NoStop}%
\bibitem [{\citenamefont {Frucci}\ \emph {et~al.}(2010)\citenamefont {Frucci}, \citenamefont {Di~Gaspare}, \citenamefont {Evangelisti}, \citenamefont {Giovine}, \citenamefont {Notargiacomo}, \citenamefont {Piazza},\ and\ \citenamefont {Beltram}}]{Frucci2010}%
  \BibitemOpen
  \bibfield  {author} {\bibinfo {author} {\bibfnamefont {G.}~\bibnamefont {Frucci}}, \bibinfo {author} {\bibfnamefont {L.}~\bibnamefont {Di~Gaspare}}, \bibinfo {author} {\bibfnamefont {F.}~\bibnamefont {Evangelisti}}, \bibinfo {author} {\bibfnamefont {E.}~\bibnamefont {Giovine}}, \bibinfo {author} {\bibfnamefont {A.}~\bibnamefont {Notargiacomo}}, \bibinfo {author} {\bibfnamefont {V.}~\bibnamefont {Piazza}},\ and\ \bibinfo {author} {\bibfnamefont {F.}~\bibnamefont {Beltram}},\ }\bibfield  {title} {\bibinfo {title} {Conductance and valley splitting in etched {Si/SiGe} one-dimensional nanostructures},\ }\href {https://doi.org/10.1103/PhysRevB.81.195311} {\bibfield  {journal} {\bibinfo  {journal} {Phys. Rev. B}\ }\textbf {\bibinfo {volume} {81}},\ \bibinfo {pages} {195311} (\bibinfo {year} {2010})}\BibitemShut {NoStop}%
\bibitem [{\citenamefont {Overweg}\ \emph {et~al.}(2018)\citenamefont {Overweg}, \citenamefont {Knothe}, \citenamefont {Fabian}, \citenamefont {Linhart}, \citenamefont {Rickhaus}, \citenamefont {Wernli}, \citenamefont {Watanabe}, \citenamefont {Taniguchi}, \citenamefont {S\'anchez}, \citenamefont {Burgd\"orfer}, \citenamefont {Libisch}, \citenamefont {Fal'ko}, \citenamefont {Ensslin},\ and\ \citenamefont {Ihn}}]{Overweg2018PRL}%
  \BibitemOpen
  \bibfield  {author} {\bibinfo {author} {\bibfnamefont {H.}~\bibnamefont {Overweg}}, \bibinfo {author} {\bibfnamefont {A.}~\bibnamefont {Knothe}}, \bibinfo {author} {\bibfnamefont {T.}~\bibnamefont {Fabian}}, \bibinfo {author} {\bibfnamefont {L.}~\bibnamefont {Linhart}}, \bibinfo {author} {\bibfnamefont {P.}~\bibnamefont {Rickhaus}}, \bibinfo {author} {\bibfnamefont {L.}~\bibnamefont {Wernli}}, \bibinfo {author} {\bibfnamefont {K.}~\bibnamefont {Watanabe}}, \bibinfo {author} {\bibfnamefont {T.}~\bibnamefont {Taniguchi}}, \bibinfo {author} {\bibfnamefont {D.}~\bibnamefont {S\'anchez}}, \bibinfo {author} {\bibfnamefont {J.}~\bibnamefont {Burgd\"orfer}}, \bibinfo {author} {\bibfnamefont {F.}~\bibnamefont {Libisch}}, \bibinfo {author} {\bibfnamefont {V.~I.}\ \bibnamefont {Fal'ko}}, \bibinfo {author} {\bibfnamefont {K.}~\bibnamefont {Ensslin}},\ and\ \bibinfo {author} {\bibfnamefont {T.}~\bibnamefont {Ihn}},\ }\bibfield  {title} {\bibinfo {title} {Topologically nontrivial valley states in bilayer graphene quantum
  point contacts},\ }\href {https://doi.org/10.1103/PhysRevLett.121.257702} {\bibfield  {journal} {\bibinfo  {journal} {Phys. Rev. Lett.}\ }\textbf {\bibinfo {volume} {121}},\ \bibinfo {pages} {257702} (\bibinfo {year} {2018})}\BibitemShut {NoStop}%
\bibitem [{\citenamefont {Kraft}\ \emph {et~al.}(2018)\citenamefont {Kraft}, \citenamefont {Krainov}, \citenamefont {Gall}, \citenamefont {Dmitriev}, \citenamefont {Krupke}, \citenamefont {Gornyi},\ and\ \citenamefont {Danneau}}]{Kraft2018}%
  \BibitemOpen
  \bibfield  {author} {\bibinfo {author} {\bibfnamefont {R.}~\bibnamefont {Kraft}}, \bibinfo {author} {\bibfnamefont {I.~V.}\ \bibnamefont {Krainov}}, \bibinfo {author} {\bibfnamefont {V.}~\bibnamefont {Gall}}, \bibinfo {author} {\bibfnamefont {A.~P.}\ \bibnamefont {Dmitriev}}, \bibinfo {author} {\bibfnamefont {R.}~\bibnamefont {Krupke}}, \bibinfo {author} {\bibfnamefont {I.~V.}\ \bibnamefont {Gornyi}},\ and\ \bibinfo {author} {\bibfnamefont {R.}~\bibnamefont {Danneau}},\ }\bibfield  {title} {\bibinfo {title} {Valley subband splitting in bilayer graphene quantum point contacts},\ }\href {https://doi.org/10.1103/PhysRevLett.121.257703} {\bibfield  {journal} {\bibinfo  {journal} {Phys. Rev. Lett.}\ }\textbf {\bibinfo {volume} {121}},\ \bibinfo {pages} {257703} (\bibinfo {year} {2018})}\BibitemShut {NoStop}%
\bibitem [{\citenamefont {Banszerus}\ \emph {et~al.}(2020)\citenamefont {Banszerus}, \citenamefont {Frohn}, \citenamefont {Fabian}, \citenamefont {Somanchi}, \citenamefont {Epping}, \citenamefont {M\"uller}, \citenamefont {Neumaier}, \citenamefont {Watanabe}, \citenamefont {Taniguchi}, \citenamefont {Libisch}, \citenamefont {Beschoten}, \citenamefont {Hassler},\ and\ \citenamefont {Stampfer}}]{Banszerus_SOC_QPC}%
  \BibitemOpen
  \bibfield  {author} {\bibinfo {author} {\bibfnamefont {L.}~\bibnamefont {Banszerus}}, \bibinfo {author} {\bibfnamefont {B.}~\bibnamefont {Frohn}}, \bibinfo {author} {\bibfnamefont {T.}~\bibnamefont {Fabian}}, \bibinfo {author} {\bibfnamefont {S.}~\bibnamefont {Somanchi}}, \bibinfo {author} {\bibfnamefont {A.}~\bibnamefont {Epping}}, \bibinfo {author} {\bibfnamefont {M.}~\bibnamefont {M\"uller}}, \bibinfo {author} {\bibfnamefont {D.}~\bibnamefont {Neumaier}}, \bibinfo {author} {\bibfnamefont {K.}~\bibnamefont {Watanabe}}, \bibinfo {author} {\bibfnamefont {T.}~\bibnamefont {Taniguchi}}, \bibinfo {author} {\bibfnamefont {F.}~\bibnamefont {Libisch}}, \bibinfo {author} {\bibfnamefont {B.}~\bibnamefont {Beschoten}}, \bibinfo {author} {\bibfnamefont {F.}~\bibnamefont {Hassler}},\ and\ \bibinfo {author} {\bibfnamefont {C.}~\bibnamefont {Stampfer}},\ }\bibfield  {title} {\bibinfo {title} {Observation of the spin-orbit gap in bilayer graphene by one-dimensional ballistic transport},\ }\href
  {https://doi.org/10.1103/PhysRevLett.124.177701} {\bibfield  {journal} {\bibinfo  {journal} {Phys. Rev. Lett.}\ }\textbf {\bibinfo {volume} {124}},\ \bibinfo {pages} {177701} (\bibinfo {year} {2020})}\BibitemShut {NoStop}%
\bibitem [{\citenamefont {Kurzmann}\ \emph {et~al.}(2021)\citenamefont {Kurzmann}, \citenamefont {Kleeorin}, \citenamefont {Tong}, \citenamefont {Garreis}, \citenamefont {Knothe}, \citenamefont {Eich}, \citenamefont {Mittag}, \citenamefont {Gold}, \citenamefont {de~Vries}, \citenamefont {Watanabe}, \citenamefont {Taniguchi}, \citenamefont {Fal'ko}, \citenamefont {Meir}, \citenamefont {Ihn},\ and\ \citenamefont {Ensslin}}]{Kurzmann2021}%
  \BibitemOpen
  \bibfield  {author} {\bibinfo {author} {\bibfnamefont {A.}~\bibnamefont {Kurzmann}}, \bibinfo {author} {\bibfnamefont {Y.}~\bibnamefont {Kleeorin}}, \bibinfo {author} {\bibfnamefont {C.}~\bibnamefont {Tong}}, \bibinfo {author} {\bibfnamefont {R.}~\bibnamefont {Garreis}}, \bibinfo {author} {\bibfnamefont {A.}~\bibnamefont {Knothe}}, \bibinfo {author} {\bibfnamefont {M.}~\bibnamefont {Eich}}, \bibinfo {author} {\bibfnamefont {C.}~\bibnamefont {Mittag}}, \bibinfo {author} {\bibfnamefont {C.}~\bibnamefont {Gold}}, \bibinfo {author} {\bibfnamefont {F.~K.}\ \bibnamefont {de~Vries}}, \bibinfo {author} {\bibfnamefont {K.}~\bibnamefont {Watanabe}}, \bibinfo {author} {\bibfnamefont {T.}~\bibnamefont {Taniguchi}}, \bibinfo {author} {\bibfnamefont {V.}~\bibnamefont {Fal'ko}}, \bibinfo {author} {\bibfnamefont {Y.}~\bibnamefont {Meir}}, \bibinfo {author} {\bibfnamefont {T.}~\bibnamefont {Ihn}},\ and\ \bibinfo {author} {\bibfnamefont {K.}~\bibnamefont {Ensslin}},\ }\bibfield  {title} {\bibinfo {title} {Kondo effect and
  spin--orbit coupling in graphene quantum dots},\ }\href {https://doi.org/10.1038/s41467-021-26149-3} {\bibfield  {journal} {\bibinfo  {journal} {Nature Communications}\ }\textbf {\bibinfo {volume} {12}},\ \bibinfo {pages} {6004} (\bibinfo {year} {2021})}\BibitemShut {NoStop}%
\bibitem [{\citenamefont {Gall}\ \emph {et~al.}(2022)\citenamefont {Gall}, \citenamefont {Kraft}, \citenamefont {Gornyi},\ and\ \citenamefont {Danneau}}]{Gall2022}%
  \BibitemOpen
  \bibfield  {author} {\bibinfo {author} {\bibfnamefont {V.}~\bibnamefont {Gall}}, \bibinfo {author} {\bibfnamefont {R.}~\bibnamefont {Kraft}}, \bibinfo {author} {\bibfnamefont {I.~V.}\ \bibnamefont {Gornyi}},\ and\ \bibinfo {author} {\bibfnamefont {R.}~\bibnamefont {Danneau}},\ }\bibfield  {title} {\bibinfo {title} {Spin and valley degrees of freedom in a bilayer graphene quantum point contact: Zeeman splitting and interaction effects},\ }\href {https://doi.org/10.1103/PhysRevResearch.4.023142} {\bibfield  {journal} {\bibinfo  {journal} {Phys. Rev. Res.}\ }\textbf {\bibinfo {volume} {4}},\ \bibinfo {pages} {023142} (\bibinfo {year} {2022})}\BibitemShut {NoStop}%
\bibitem [{\citenamefont {B\"uttiker}(1990)}]{Buttiker1990}%
  \BibitemOpen
  \bibfield  {author} {\bibinfo {author} {\bibfnamefont {M.}~\bibnamefont {B\"uttiker}},\ }\bibfield  {title} {\bibinfo {title} {Quantized transmission of a saddle-point constriction},\ }\href {https://doi.org/10.1103/PhysRevB.41.7906} {\bibfield  {journal} {\bibinfo  {journal} {Phys. Rev. B}\ }\textbf {\bibinfo {volume} {41}},\ \bibinfo {pages} {7906} (\bibinfo {year} {1990})}\BibitemShut {NoStop}%
\bibitem [{\citenamefont {Gmitra}\ and\ \citenamefont {Fabian}(2017)}]{Gmitra2017}%
  \BibitemOpen
  \bibfield  {author} {\bibinfo {author} {\bibfnamefont {M.}~\bibnamefont {Gmitra}}\ and\ \bibinfo {author} {\bibfnamefont {J.}~\bibnamefont {Fabian}},\ }\bibfield  {title} {\bibinfo {title} {Proximity effects in bilayer graphene on monolayer {${\mathrm{WSe}}_{2}$}: Field-effect spin valley locking, spin-orbit valve, and spin transistor},\ }\href {https://doi.org/10.1103/PhysRevLett.119.146401} {\bibfield  {journal} {\bibinfo  {journal} {Phys. Rev. Lett.}\ }\textbf {\bibinfo {volume} {119}},\ \bibinfo {pages} {146401} (\bibinfo {year} {2017})}\BibitemShut {NoStop}%
\bibitem [{\citenamefont {Avsar}\ \emph {et~al.}(2014)\citenamefont {Avsar}, \citenamefont {Tan}, \citenamefont {Taychatanapat}, \citenamefont {Balakrishnan}, \citenamefont {Koon}, \citenamefont {Yeo}, \citenamefont {Lahiri}, \citenamefont {Carvalho}, \citenamefont {Rodin}, \citenamefont {O'Farrell}, \citenamefont {Eda}, \citenamefont {Castro~Neto},\ and\ \citenamefont {{\"O}zyilmaz}}]{Avsar2014}%
  \BibitemOpen
  \bibfield  {author} {\bibinfo {author} {\bibfnamefont {A.}~\bibnamefont {Avsar}}, \bibinfo {author} {\bibfnamefont {J.~Y.}\ \bibnamefont {Tan}}, \bibinfo {author} {\bibfnamefont {T.}~\bibnamefont {Taychatanapat}}, \bibinfo {author} {\bibfnamefont {J.}~\bibnamefont {Balakrishnan}}, \bibinfo {author} {\bibfnamefont {G.~K.~W.}\ \bibnamefont {Koon}}, \bibinfo {author} {\bibfnamefont {Y.}~\bibnamefont {Yeo}}, \bibinfo {author} {\bibfnamefont {J.}~\bibnamefont {Lahiri}}, \bibinfo {author} {\bibfnamefont {A.}~\bibnamefont {Carvalho}}, \bibinfo {author} {\bibfnamefont {A.~S.}\ \bibnamefont {Rodin}}, \bibinfo {author} {\bibfnamefont {E.~C.~T.}\ \bibnamefont {O'Farrell}}, \bibinfo {author} {\bibfnamefont {G.}~\bibnamefont {Eda}}, \bibinfo {author} {\bibfnamefont {A.~H.}\ \bibnamefont {Castro~Neto}},\ and\ \bibinfo {author} {\bibfnamefont {B.}~\bibnamefont {{\"O}zyilmaz}},\ }\bibfield  {title} {\bibinfo {title} {Spin--orbit proximity effect in graphene},\ }\href {https://doi.org/10.1038/ncomms5875} {\bibfield
  {journal} {\bibinfo  {journal} {Nature Communications}\ }\textbf {\bibinfo {volume} {5}},\ \bibinfo {pages} {4875} (\bibinfo {year} {2014})}\BibitemShut {NoStop}%
\bibitem [{\citenamefont {Wang}\ \emph {et~al.}(2015)\citenamefont {Wang}, \citenamefont {Ki}, \citenamefont {Chen}, \citenamefont {Berger}, \citenamefont {MacDonald},\ and\ \citenamefont {Morpurgo}}]{Wang2015}%
  \BibitemOpen
  \bibfield  {author} {\bibinfo {author} {\bibfnamefont {Z.}~\bibnamefont {Wang}}, \bibinfo {author} {\bibfnamefont {D.-K.}\ \bibnamefont {Ki}}, \bibinfo {author} {\bibfnamefont {H.}~\bibnamefont {Chen}}, \bibinfo {author} {\bibfnamefont {H.}~\bibnamefont {Berger}}, \bibinfo {author} {\bibfnamefont {A.~H.}\ \bibnamefont {MacDonald}},\ and\ \bibinfo {author} {\bibfnamefont {A.~F.}\ \bibnamefont {Morpurgo}},\ }\bibfield  {title} {\bibinfo {title} {Strong interface-induced spin--orbit interaction in graphene on {${\mathrm{WS}}_{2}$}},\ }\href {https://doi.org/10.1038/ncomms9339} {\bibfield  {journal} {\bibinfo  {journal} {Nature Communications}\ }\textbf {\bibinfo {volume} {6}},\ \bibinfo {pages} {8339} (\bibinfo {year} {2015})}\BibitemShut {NoStop}%
\bibitem [{\citenamefont {Yang}\ \emph {et~al.}(2016)\citenamefont {Yang}, \citenamefont {Tu}, \citenamefont {Kim}, \citenamefont {Wu}, \citenamefont {Wang}, \citenamefont {Alicea}, \citenamefont {Wu}, \citenamefont {Bockrath},\ and\ \citenamefont {Shi}}]{Yang_2016}%
  \BibitemOpen
  \bibfield  {author} {\bibinfo {author} {\bibfnamefont {B.}~\bibnamefont {Yang}}, \bibinfo {author} {\bibfnamefont {M.-F.}\ \bibnamefont {Tu}}, \bibinfo {author} {\bibfnamefont {J.}~\bibnamefont {Kim}}, \bibinfo {author} {\bibfnamefont {Y.}~\bibnamefont {Wu}}, \bibinfo {author} {\bibfnamefont {H.}~\bibnamefont {Wang}}, \bibinfo {author} {\bibfnamefont {J.}~\bibnamefont {Alicea}}, \bibinfo {author} {\bibfnamefont {R.}~\bibnamefont {Wu}}, \bibinfo {author} {\bibfnamefont {M.}~\bibnamefont {Bockrath}},\ and\ \bibinfo {author} {\bibfnamefont {J.}~\bibnamefont {Shi}},\ }\bibfield  {title} {\bibinfo {title} {Tunable spin–orbit coupling and symmetry-protected edge states in graphene/{${\mathrm{WS}}_{2}$}},\ }\href {https://doi.org/10.1088/2053-1583/3/3/031012} {\bibfield  {journal} {\bibinfo  {journal} {2D Materials}\ }\textbf {\bibinfo {volume} {3}},\ \bibinfo {pages} {031012} (\bibinfo {year} {2016})}\BibitemShut {NoStop}%
\bibitem [{\citenamefont {Gerber}\ \emph {et~al.}(2025)\citenamefont {Gerber}, \citenamefont {Ersoy}, \citenamefont {Masseroni}, \citenamefont {Niese}, \citenamefont {Laumer}, \citenamefont {Denisov}, \citenamefont {Duprez}, \citenamefont {Huang}, \citenamefont {Adam}, \citenamefont {Ostertag}, \citenamefont {Tong}, \citenamefont {Taniguchi}, \citenamefont {Watanabe}, \citenamefont {Fal'ko}, \citenamefont {Ihn}, \citenamefont {Ensslin},\ and\ \citenamefont {Knothe}}]{Gerber2025}%
  \BibitemOpen
  \bibfield  {author} {\bibinfo {author} {\bibfnamefont {J.~D.}\ \bibnamefont {Gerber}}, \bibinfo {author} {\bibfnamefont {E.}~\bibnamefont {Ersoy}}, \bibinfo {author} {\bibfnamefont {M.}~\bibnamefont {Masseroni}}, \bibinfo {author} {\bibfnamefont {M.}~\bibnamefont {Niese}}, \bibinfo {author} {\bibfnamefont {M.}~\bibnamefont {Laumer}}, \bibinfo {author} {\bibfnamefont {A.~O.}\ \bibnamefont {Denisov}}, \bibinfo {author} {\bibfnamefont {H.}~\bibnamefont {Duprez}}, \bibinfo {author} {\bibfnamefont {W.~W.}\ \bibnamefont {Huang}}, \bibinfo {author} {\bibfnamefont {C.}~\bibnamefont {Adam}}, \bibinfo {author} {\bibfnamefont {L.}~\bibnamefont {Ostertag}}, \bibinfo {author} {\bibfnamefont {C.}~\bibnamefont {Tong}}, \bibinfo {author} {\bibfnamefont {T.}~\bibnamefont {Taniguchi}}, \bibinfo {author} {\bibfnamefont {K.}~\bibnamefont {Watanabe}}, \bibinfo {author} {\bibfnamefont {V.~I.}\ \bibnamefont {Fal'ko}}, \bibinfo {author} {\bibfnamefont {T.}~\bibnamefont {Ihn}}, \bibinfo {author} {\bibfnamefont {K.}~\bibnamefont
  {Ensslin}},\ and\ \bibinfo {author} {\bibfnamefont {A.}~\bibnamefont {Knothe}},\ }\bibfield  {title} {\bibinfo {title} {Tunable spin--orbit splitting in bilayer graphene/{${\mathrm{WSe}}_{2}$} quantum devices},\ }\href {https://doi.org/10.1021/acs.nanolett.5c02309} {\bibfield  {journal} {\bibinfo  {journal} {Nano Letters}\ }\textbf {\bibinfo {volume} {25}},\ \bibinfo {pages} {12480} (\bibinfo {year} {2025})}\BibitemShut {NoStop}%
\bibitem [{sup()}]{supplemental2025}%
  \BibitemOpen
  \href@noop {} {}\bibinfo {note} {See Supplemental Material at URL for details on device fabrication, measurement setup, data analysis, and additional measurement data, which includes Refs. [37-39].}\BibitemShut {Stop}%
\bibitem [{\citenamefont {Lee}\ \emph {et~al.}(2020)\citenamefont {Lee}, \citenamefont {Knothe}, \citenamefont {Overweg}, \citenamefont {Eich}, \citenamefont {Gold}, \citenamefont {Kurzmann}, \citenamefont {Klasovika}, \citenamefont {Taniguchi}, \citenamefont {Wantanabe}, \citenamefont {Fal'ko}, \citenamefont {Ihn}, \citenamefont {Ensslin},\ and\ \citenamefont {Rickhaus}}]{Lee2020}%
  \BibitemOpen
  \bibfield  {author} {\bibinfo {author} {\bibfnamefont {Y.}~\bibnamefont {Lee}}, \bibinfo {author} {\bibfnamefont {A.}~\bibnamefont {Knothe}}, \bibinfo {author} {\bibfnamefont {H.}~\bibnamefont {Overweg}}, \bibinfo {author} {\bibfnamefont {M.}~\bibnamefont {Eich}}, \bibinfo {author} {\bibfnamefont {C.}~\bibnamefont {Gold}}, \bibinfo {author} {\bibfnamefont {A.}~\bibnamefont {Kurzmann}}, \bibinfo {author} {\bibfnamefont {V.}~\bibnamefont {Klasovika}}, \bibinfo {author} {\bibfnamefont {T.}~\bibnamefont {Taniguchi}}, \bibinfo {author} {\bibfnamefont {K.}~\bibnamefont {Wantanabe}}, \bibinfo {author} {\bibfnamefont {V.}~\bibnamefont {Fal'ko}}, \bibinfo {author} {\bibfnamefont {T.}~\bibnamefont {Ihn}}, \bibinfo {author} {\bibfnamefont {K.}~\bibnamefont {Ensslin}},\ and\ \bibinfo {author} {\bibfnamefont {P.}~\bibnamefont {Rickhaus}},\ }\bibfield  {title} {\bibinfo {title} {Tunable valley splitting due to topological orbital magnetic moment in bilayer graphene quantum point contacts},\ }\href
  {https://doi.org/10.1103/PhysRevLett.124.126802} {\bibfield  {journal} {\bibinfo  {journal} {Phys. Rev. Lett.}\ }\textbf {\bibinfo {volume} {124}},\ \bibinfo {pages} {126802} (\bibinfo {year} {2020})}\BibitemShut {NoStop}%
\bibitem [{\citenamefont {Banszerus}\ \emph {et~al.}(2021)\citenamefont {Banszerus}, \citenamefont {M{\"o}ller}, \citenamefont {Steiner}, \citenamefont {Icking}, \citenamefont {Trellenkamp}, \citenamefont {Lentz}, \citenamefont {Watanabe}, \citenamefont {Taniguchi}, \citenamefont {Volk},\ and\ \citenamefont {Stampfer}}]{Banszerus2021_SpinValley}%
  \BibitemOpen
  \bibfield  {author} {\bibinfo {author} {\bibfnamefont {L.}~\bibnamefont {Banszerus}}, \bibinfo {author} {\bibfnamefont {S.}~\bibnamefont {M{\"o}ller}}, \bibinfo {author} {\bibfnamefont {C.}~\bibnamefont {Steiner}}, \bibinfo {author} {\bibfnamefont {E.}~\bibnamefont {Icking}}, \bibinfo {author} {\bibfnamefont {S.}~\bibnamefont {Trellenkamp}}, \bibinfo {author} {\bibfnamefont {F.}~\bibnamefont {Lentz}}, \bibinfo {author} {\bibfnamefont {K.}~\bibnamefont {Watanabe}}, \bibinfo {author} {\bibfnamefont {T.}~\bibnamefont {Taniguchi}}, \bibinfo {author} {\bibfnamefont {C.}~\bibnamefont {Volk}},\ and\ \bibinfo {author} {\bibfnamefont {C.}~\bibnamefont {Stampfer}},\ }\bibfield  {title} {\bibinfo {title} {Spin-valley coupling in single-electron bilayer graphene quantum dots},\ }\href {https://doi.org/10.1038/s41467-021-25498-3} {\bibfield  {journal} {\bibinfo  {journal} {Nature Communications}\ }\textbf {\bibinfo {volume} {12}},\ \bibinfo {pages} {5250} (\bibinfo {year} {2021})}\BibitemShut {NoStop}%
\bibitem [{\citenamefont {Smith}\ \emph {et~al.}(2016)\citenamefont {Smith}, \citenamefont {Al-Taie}, \citenamefont {Lesage}, \citenamefont {Thomas}, \citenamefont {Sfigakis}, \citenamefont {See}, \citenamefont {Griffiths}, \citenamefont {Farrer}, \citenamefont {Jones}, \citenamefont {Ritchie}, \citenamefont {Kelly},\ and\ \citenamefont {Smith}}]{Smith2016}%
  \BibitemOpen
  \bibfield  {author} {\bibinfo {author} {\bibfnamefont {L.~W.}\ \bibnamefont {Smith}}, \bibinfo {author} {\bibfnamefont {H.}~\bibnamefont {Al-Taie}}, \bibinfo {author} {\bibfnamefont {A.~A.~J.}\ \bibnamefont {Lesage}}, \bibinfo {author} {\bibfnamefont {K.~J.}\ \bibnamefont {Thomas}}, \bibinfo {author} {\bibfnamefont {F.}~\bibnamefont {Sfigakis}}, \bibinfo {author} {\bibfnamefont {P.}~\bibnamefont {See}}, \bibinfo {author} {\bibfnamefont {J.~P.}\ \bibnamefont {Griffiths}}, \bibinfo {author} {\bibfnamefont {I.}~\bibnamefont {Farrer}}, \bibinfo {author} {\bibfnamefont {G.~A.~C.}\ \bibnamefont {Jones}}, \bibinfo {author} {\bibfnamefont {D.~A.}\ \bibnamefont {Ritchie}}, \bibinfo {author} {\bibfnamefont {M.~J.}\ \bibnamefont {Kelly}},\ and\ \bibinfo {author} {\bibfnamefont {C.~G.}\ \bibnamefont {Smith}},\ }\bibfield  {title} {\bibinfo {title} {Effect of split gate size on the electrostatic potential and 0.7 anomaly within quantum wires on a modulation-doped $\mathrm{GaAs}/\mathrm{AlGaAs}$ heterostructure},\ }\href
  {https://doi.org/10.1103/PhysRevApplied.5.044015} {\bibfield  {journal} {\bibinfo  {journal} {Phys. Rev. Appl.}\ }\textbf {\bibinfo {volume} {5}},\ \bibinfo {pages} {044015} (\bibinfo {year} {2016})}\BibitemShut {NoStop}%
\bibitem [{\citenamefont {Graham}\ \emph {et~al.}(2003)\citenamefont {Graham}, \citenamefont {Thomas}, \citenamefont {Pepper}, \citenamefont {Cooper}, \citenamefont {Simmons},\ and\ \citenamefont {Ritchie}}]{Graham2003}%
  \BibitemOpen
  \bibfield  {author} {\bibinfo {author} {\bibfnamefont {A.~C.}\ \bibnamefont {Graham}}, \bibinfo {author} {\bibfnamefont {K.~J.}\ \bibnamefont {Thomas}}, \bibinfo {author} {\bibfnamefont {M.}~\bibnamefont {Pepper}}, \bibinfo {author} {\bibfnamefont {N.~R.}\ \bibnamefont {Cooper}}, \bibinfo {author} {\bibfnamefont {M.~Y.}\ \bibnamefont {Simmons}},\ and\ \bibinfo {author} {\bibfnamefont {D.~A.}\ \bibnamefont {Ritchie}},\ }\bibfield  {title} {\bibinfo {title} {Interaction effects at crossings of spin-polarized one-dimensional subbands},\ }\href {https://doi.org/10.1103/PhysRevLett.91.136404} {\bibfield  {journal} {\bibinfo  {journal} {Phys. Rev. Lett.}\ }\textbf {\bibinfo {volume} {91}},\ \bibinfo {pages} {136404} (\bibinfo {year} {2003})}\BibitemShut {NoStop}%
\bibitem [{\citenamefont {Knothe}\ and\ \citenamefont {Fal'ko}(2020)}]{Knothe2020}%
  \BibitemOpen
  \bibfield  {author} {\bibinfo {author} {\bibfnamefont {A.}~\bibnamefont {Knothe}}\ and\ \bibinfo {author} {\bibfnamefont {V.}~\bibnamefont {Fal'ko}},\ }\bibfield  {title} {\bibinfo {title} {Quartet states in two-electron quantum dots in bilayer graphene},\ }\href {https://doi.org/10.1103/PhysRevB.101.235423} {\bibfield  {journal} {\bibinfo  {journal} {Phys. Rev. B}\ }\textbf {\bibinfo {volume} {101}},\ \bibinfo {pages} {235423} (\bibinfo {year} {2020})}\BibitemShut {NoStop}%
\bibitem [{\citenamefont {Seiler}\ \emph {et~al.}(2022)\citenamefont {Seiler}, \citenamefont {Geisenhof}, \citenamefont {Winterer}, \citenamefont {Watanabe}, \citenamefont {Taniguchi}, \citenamefont {Xu}, \citenamefont {Zhang},\ and\ \citenamefont {Weitz}}]{Seiler2022}%
  \BibitemOpen
  \bibfield  {author} {\bibinfo {author} {\bibfnamefont {A.~M.}\ \bibnamefont {Seiler}}, \bibinfo {author} {\bibfnamefont {F.~R.}\ \bibnamefont {Geisenhof}}, \bibinfo {author} {\bibfnamefont {F.}~\bibnamefont {Winterer}}, \bibinfo {author} {\bibfnamefont {K.}~\bibnamefont {Watanabe}}, \bibinfo {author} {\bibfnamefont {T.}~\bibnamefont {Taniguchi}}, \bibinfo {author} {\bibfnamefont {T.}~\bibnamefont {Xu}}, \bibinfo {author} {\bibfnamefont {F.}~\bibnamefont {Zhang}},\ and\ \bibinfo {author} {\bibfnamefont {R.~T.}\ \bibnamefont {Weitz}},\ }\bibfield  {title} {\bibinfo {title} {Quantum cascade of correlated phases in trigonally warped bilayer graphene},\ }\href {https://doi.org/10.1038/s41586-022-04937-1} {\bibfield  {journal} {\bibinfo  {journal} {Nature}\ }\textbf {\bibinfo {volume} {608}},\ \bibinfo {pages} {298} (\bibinfo {year} {2022})}\BibitemShut {NoStop}%
\bibitem [{\citenamefont {Zhou}\ \emph {et~al.}(2022)\citenamefont {Zhou}, \citenamefont {Holleis}, \citenamefont {Saito}, \citenamefont {Cohen}, \citenamefont {Huynh}, \citenamefont {Patterson}, \citenamefont {Yang}, \citenamefont {Taniguchi}, \citenamefont {Watanabe},\ and\ \citenamefont {Young}}]{Zhou2022}%
  \BibitemOpen
  \bibfield  {author} {\bibinfo {author} {\bibfnamefont {H.}~\bibnamefont {Zhou}}, \bibinfo {author} {\bibfnamefont {L.}~\bibnamefont {Holleis}}, \bibinfo {author} {\bibfnamefont {Y.}~\bibnamefont {Saito}}, \bibinfo {author} {\bibfnamefont {L.}~\bibnamefont {Cohen}}, \bibinfo {author} {\bibfnamefont {W.}~\bibnamefont {Huynh}}, \bibinfo {author} {\bibfnamefont {C.~L.}\ \bibnamefont {Patterson}}, \bibinfo {author} {\bibfnamefont {F.}~\bibnamefont {Yang}}, \bibinfo {author} {\bibfnamefont {T.}~\bibnamefont {Taniguchi}}, \bibinfo {author} {\bibfnamefont {K.}~\bibnamefont {Watanabe}},\ and\ \bibinfo {author} {\bibfnamefont {A.~F.}\ \bibnamefont {Young}},\ }\bibfield  {title} {\bibinfo {title} {Isospin magnetism and spin-polarized superconductivity in bernal bilayer graphene},\ }\href {https://doi.org/10.1126/science.abm8386} {\bibfield  {journal} {\bibinfo  {journal} {Science}\ }\textbf {\bibinfo {volume} {375}},\ \bibinfo {pages} {774} (\bibinfo {year} {2022})}\BibitemShut {NoStop}%
\bibitem [{\citenamefont {de~la Barrera}\ \emph {et~al.}(2022)\citenamefont {de~la Barrera}, \citenamefont {Aronson}, \citenamefont {Zheng}, \citenamefont {Watanabe}, \citenamefont {Taniguchi}, \citenamefont {Ma}, \citenamefont {Jarillo-Herrero},\ and\ \citenamefont {Ashoori}}]{delaBarrera2022}%
  \BibitemOpen
  \bibfield  {author} {\bibinfo {author} {\bibfnamefont {S.~C.}\ \bibnamefont {de~la Barrera}}, \bibinfo {author} {\bibfnamefont {S.}~\bibnamefont {Aronson}}, \bibinfo {author} {\bibfnamefont {Z.}~\bibnamefont {Zheng}}, \bibinfo {author} {\bibfnamefont {K.}~\bibnamefont {Watanabe}}, \bibinfo {author} {\bibfnamefont {T.}~\bibnamefont {Taniguchi}}, \bibinfo {author} {\bibfnamefont {Q.}~\bibnamefont {Ma}}, \bibinfo {author} {\bibfnamefont {P.}~\bibnamefont {Jarillo-Herrero}},\ and\ \bibinfo {author} {\bibfnamefont {R.}~\bibnamefont {Ashoori}},\ }\bibfield  {title} {\bibinfo {title} {Cascade of isospin phase transitions in bernal-stacked bilayer graphene at zero magnetic field},\ }\href {https://doi.org/10.1038/s41567-022-01616-w} {\bibfield  {journal} {\bibinfo  {journal} {Nature Physics}\ }\textbf {\bibinfo {volume} {18}},\ \bibinfo {pages} {771} (\bibinfo {year} {2022})}\BibitemShut {NoStop}%
\bibitem [{\citenamefont {Rozhansky}\ and\ \citenamefont {Fal'ko}(2024)}]{Rozhansky2024}%
  \BibitemOpen
  \bibfield  {author} {\bibinfo {author} {\bibfnamefont {I.}~\bibnamefont {Rozhansky}}\ and\ \bibinfo {author} {\bibfnamefont {V.}~\bibnamefont {Fal'ko}},\ }\bibfield  {title} {\bibinfo {title} {Exchange-enhanced spin-orbit splitting and its density dependence for electrons in monolayer transition metal dichalcogenides},\ }\href {https://doi.org/10.1103/PhysRevB.110.L161404} {\bibfield  {journal} {\bibinfo  {journal} {Phys. Rev. B}\ }\textbf {\bibinfo {volume} {110}},\ \bibinfo {pages} {L161404} (\bibinfo {year} {2024})}\BibitemShut {NoStop}%
\bibitem [{\citenamefont {Hew}\ \emph {et~al.}(2009)\citenamefont {Hew}, \citenamefont {Thomas}, \citenamefont {Pepper}, \citenamefont {Farrer}, \citenamefont {Anderson}, \citenamefont {Jones},\ and\ \citenamefont {Ritchie}}]{Hew2009}%
  \BibitemOpen
  \bibfield  {author} {\bibinfo {author} {\bibfnamefont {W.~K.}\ \bibnamefont {Hew}}, \bibinfo {author} {\bibfnamefont {K.~J.}\ \bibnamefont {Thomas}}, \bibinfo {author} {\bibfnamefont {M.}~\bibnamefont {Pepper}}, \bibinfo {author} {\bibfnamefont {I.}~\bibnamefont {Farrer}}, \bibinfo {author} {\bibfnamefont {D.}~\bibnamefont {Anderson}}, \bibinfo {author} {\bibfnamefont {G.~A.~C.}\ \bibnamefont {Jones}},\ and\ \bibinfo {author} {\bibfnamefont {D.~A.}\ \bibnamefont {Ritchie}},\ }\bibfield  {title} {\bibinfo {title} {Incipient formation of an electron lattice in a weakly confined quantum wire},\ }\href {https://doi.org/10.1103/PhysRevLett.102.056804} {\bibfield  {journal} {\bibinfo  {journal} {Phys. Rev. Lett.}\ }\textbf {\bibinfo {volume} {102}},\ \bibinfo {pages} {056804} (\bibinfo {year} {2009})}\BibitemShut {NoStop}%
\bibitem [{\citenamefont {Kumar}\ \emph {et~al.}(2021)\citenamefont {Kumar}, \citenamefont {Pepper}, \citenamefont {Montagu}, \citenamefont {Ritchie}, \citenamefont {Farrer}, \citenamefont {Griffiths},\ and\ \citenamefont {Jones}}]{Kumar2012}%
  \BibitemOpen
  \bibfield  {author} {\bibinfo {author} {\bibfnamefont {S.}~\bibnamefont {Kumar}}, \bibinfo {author} {\bibfnamefont {M.}~\bibnamefont {Pepper}}, \bibinfo {author} {\bibfnamefont {H.}~\bibnamefont {Montagu}}, \bibinfo {author} {\bibfnamefont {D.}~\bibnamefont {Ritchie}}, \bibinfo {author} {\bibfnamefont {I.}~\bibnamefont {Farrer}}, \bibinfo {author} {\bibfnamefont {J.}~\bibnamefont {Griffiths}},\ and\ \bibinfo {author} {\bibfnamefont {G.}~\bibnamefont {Jones}},\ }\bibfield  {title} {\bibinfo {title} {Engineering electron wavefunctions in asymmetrically confined quasi one-dimensional structures},\ }\href {https://doi.org/10.1063/5.0045702} {\bibfield  {journal} {\bibinfo  {journal} {Applied Physics Letters}\ }\textbf {\bibinfo {volume} {118}},\ \bibinfo {pages} {124002} (\bibinfo {year} {2021})}\BibitemShut {NoStop}%
\end{thebibliography}


%

\clearpage
\newpage
\appendix
\renewcommand\thefigure{\thesection.\arabic{figure}}  
\renewcommand\thetable{\thesection.\arabic{table}} 

\setcounter{figure}{0}    
\setcounter{table}{0}

\onecolumngrid
\begin{center}
    \LARGE \textbf{Supplementary Material}\\[0.5cm] 

 \large{\textbf{Spin–Valley 0.7 Anomaly in Proximitized Bilayer Graphene Quantum Point Contacts}}\\[0.5cm]

    \normalsize Jonas D. Gerber\textsuperscript{1,*}, Efe Ersoy\textsuperscript{1}, Michele Masseroni\textsuperscript{1}, Markus Niese\textsuperscript{1}, Artem O. Denisov\textsuperscript{1}, Christoph Adam\textsuperscript{1}, Lara Ostertag\textsuperscript{1}, Takashi Taniguchi\textsuperscript{2}, Kenji Watanabe\textsuperscript{3}, Yigal Meir\textsuperscript{4}, Thomas Ihn\textsuperscript{1}, Klaus Ensslin\textsuperscript{1}\\[0.2cm]
    \footnotesize
    \textit{\textsuperscript{1}Solid State Physics Laboratory, ETH Zürich, 8093 Zürich, Switzerland\\
    \textsuperscript{2}Research Center for Materials Nanoarchitectonics, National Institute for Materials Science,  1-1 Namiki, Tsukuba 305-0044, Japan\\
    \textsuperscript{3}Research Center for Electronic and Optical Materials, National Institute for Materials Science, 1-1 Namiki, Tsukuba 305-0044, Japan\\
    \textsuperscript{4} Department of Physics, Ben-Gurion University of the Negev, Beer-Sheva 84105, Israel}\\
    *Email: gerberjo@phys.ethz.ch
\end{center}

\twocolumngrid

\section{Extended Data of the strong SOC sample}

Finally, we present additional data supporting the measurements in the main text.

\begin{figure}
	\includegraphics{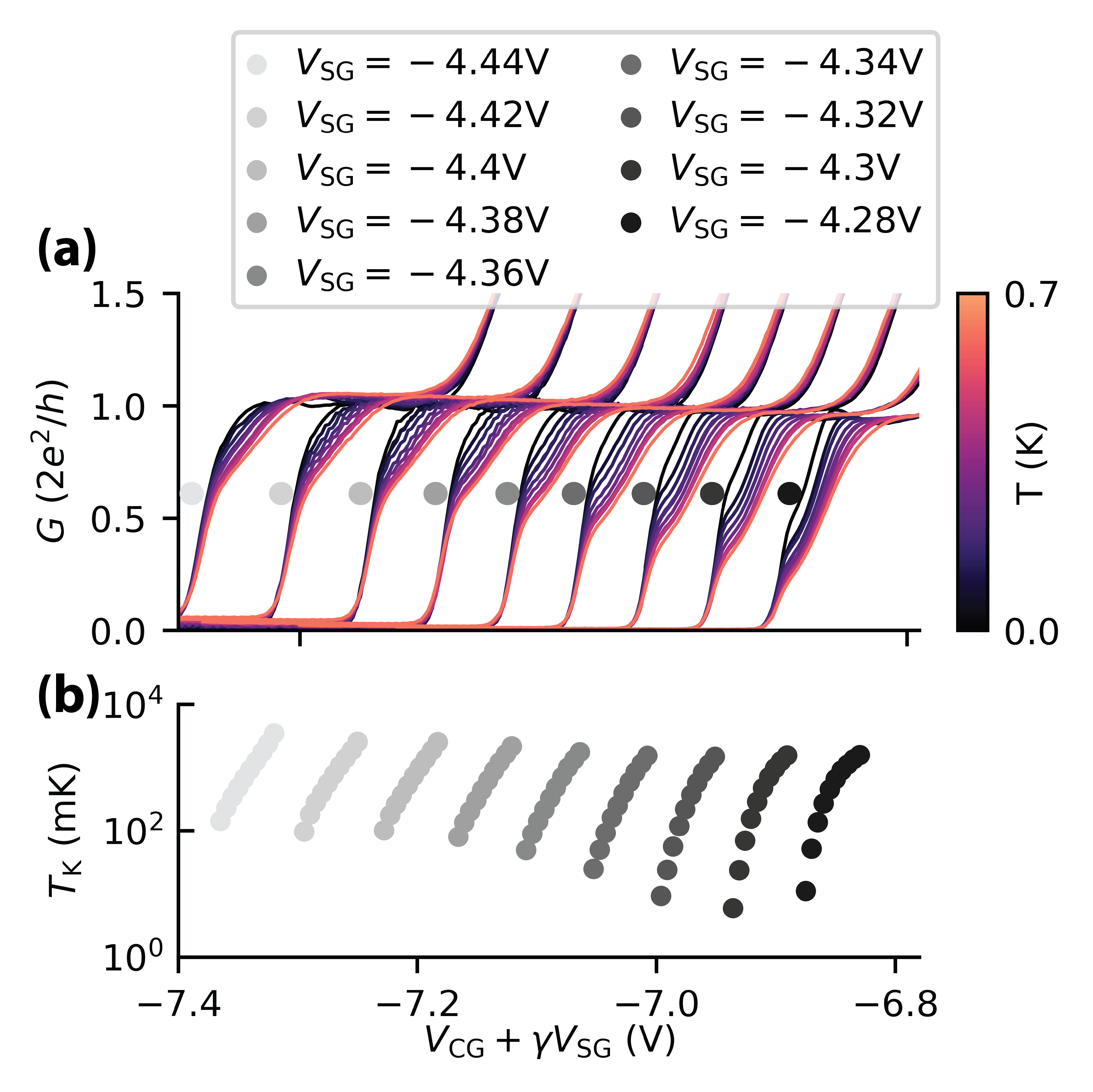}
	\caption{\textbf{(a)} Temperature-dependent conductance $G$ versus $V_\mathrm{CG}$ for varying split gate voltages $V_\mathrm{SG}$. All conductance traces show suppression at elevated temperatures typical of the 0.7 anomaly. \textbf{(b)} Respective extracted Kondo temperatures for all SG configurations. }
    	\label{FigSGTemp}
\end{figure}

Figure~\ref{FigSGTemp}a shows conductance traces as a function of split gate voltage, which tunes both $\Delta E_{1,2}$ and $\hbar\omega_x$. The 0.7 anomaly characteristics vary between split-gate configurations, including changes in the plateau height. Critically, the characteristic temperature dependence of the 0.7 anomaly persists across all configurations, confirming that the high-quality 0.7 anomaly is maintained throughout the accessible parameter space. Figure~\ref{FigSGTemp}b shows the extracted Kondo temperatures for these conductance traces. At low $V_\mathrm{SG}$ (weak confinement), Kondo temperatures drop significantly, consistent with the observation that the conductance shoulder is more suppressed at base temperature.

\begin{figure}
	\includegraphics{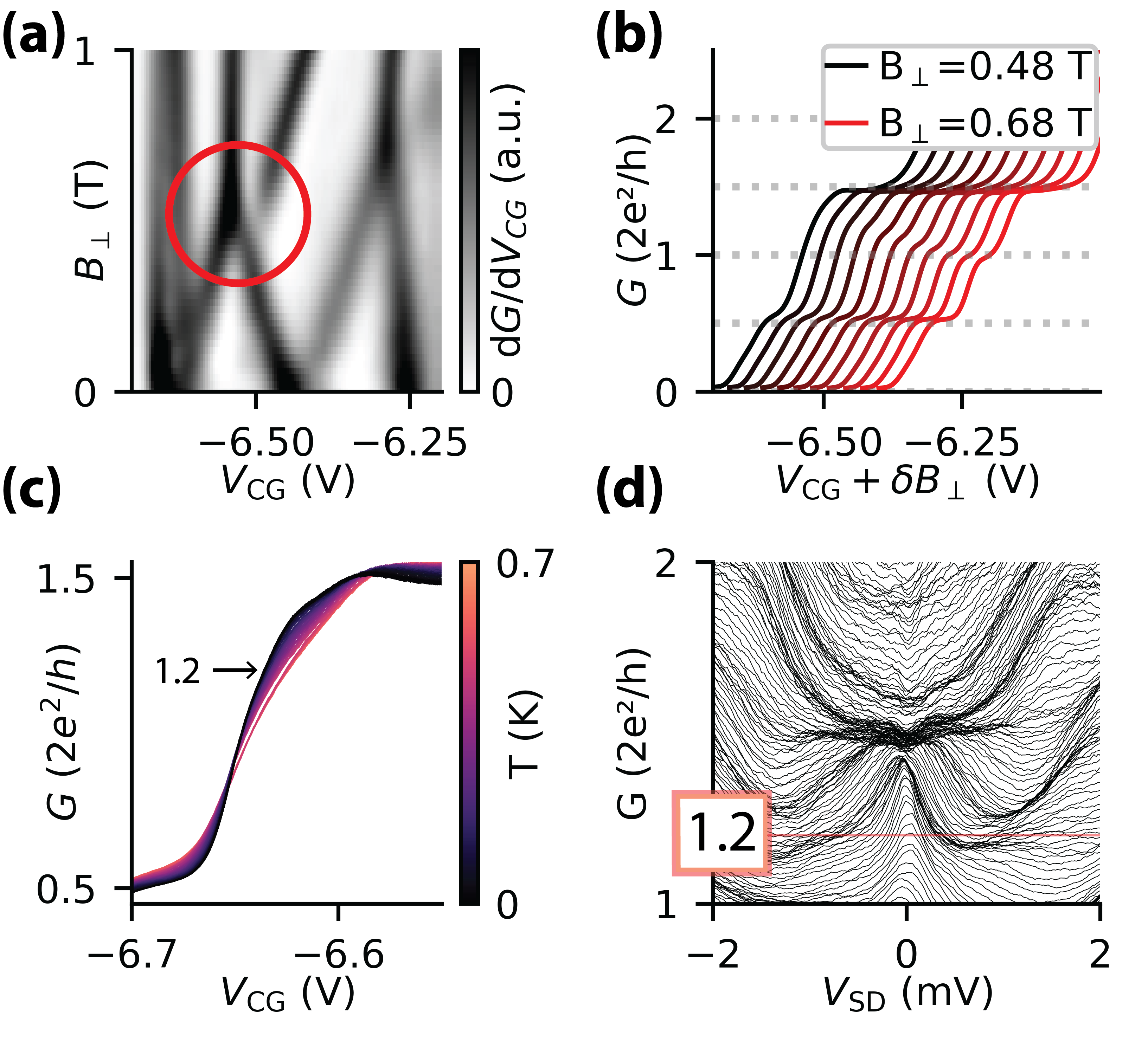}
	\caption{\textbf{(a)}  $dG/dV_\mathrm{CG}$ versus $B_\perp$ and channel gate voltage $V_\mathrm{CG}$ at \SI{10}{mK} with additional anomaly at $G=1.2G_0$ (marked by the red circle). \textbf{(b)} Respective conductance evolution in $B_\perp$. \textbf{(c)} Temperature dependence of the conductance showing $G$ suppression at elevated temperatures. \textbf{(d)} Differential conductance versus source-drain bias $V_\mathrm{SD}$ showing both a zero-bias anomaly and finite-bias plateaus at $1.2G_0$.   }
    	\label{Fig_0p7anaolg}
\end{figure}

We further identify a 0.7 analog at the crossing of the $1\mathrm{K}^-\downarrow$ and $2\mathrm{K}^+\uparrow$ modes at around $B_\perp = \SI{0.4}{T}$ (Fig.~\ref{Fig_0p7anaolg}a). The additional anomaly at $G = 1.2G_0$ evolves into a quantized $1G_0$ plateau with increasing perpendicular magnetic field (Fig.~\ref{Fig_0p7anaolg}b) and exhibits a characteristic temperature dependence (Fig.~\ref{Fig_0p7anaolg}c). We measure both a zero-bias anomaly and finite-bias plateaus at $G = 1.2G_0$ (Fig.~\ref{Fig_0p7anaolg}d), signatures consistent with the 0.7 analog previously reported in GaAs \cite{Graham2003}. Our BLG platform thus reproduces all the interaction-driven conductance features characteristic of high-quality GaAs QPCs, confirming both the excellent sample quality and the universal nature of these anomalies.

\clearpage
\newpage
\section{0.7 anomaly indications in a weak SOC sample}

\begin{figure}[h]
	\includegraphics{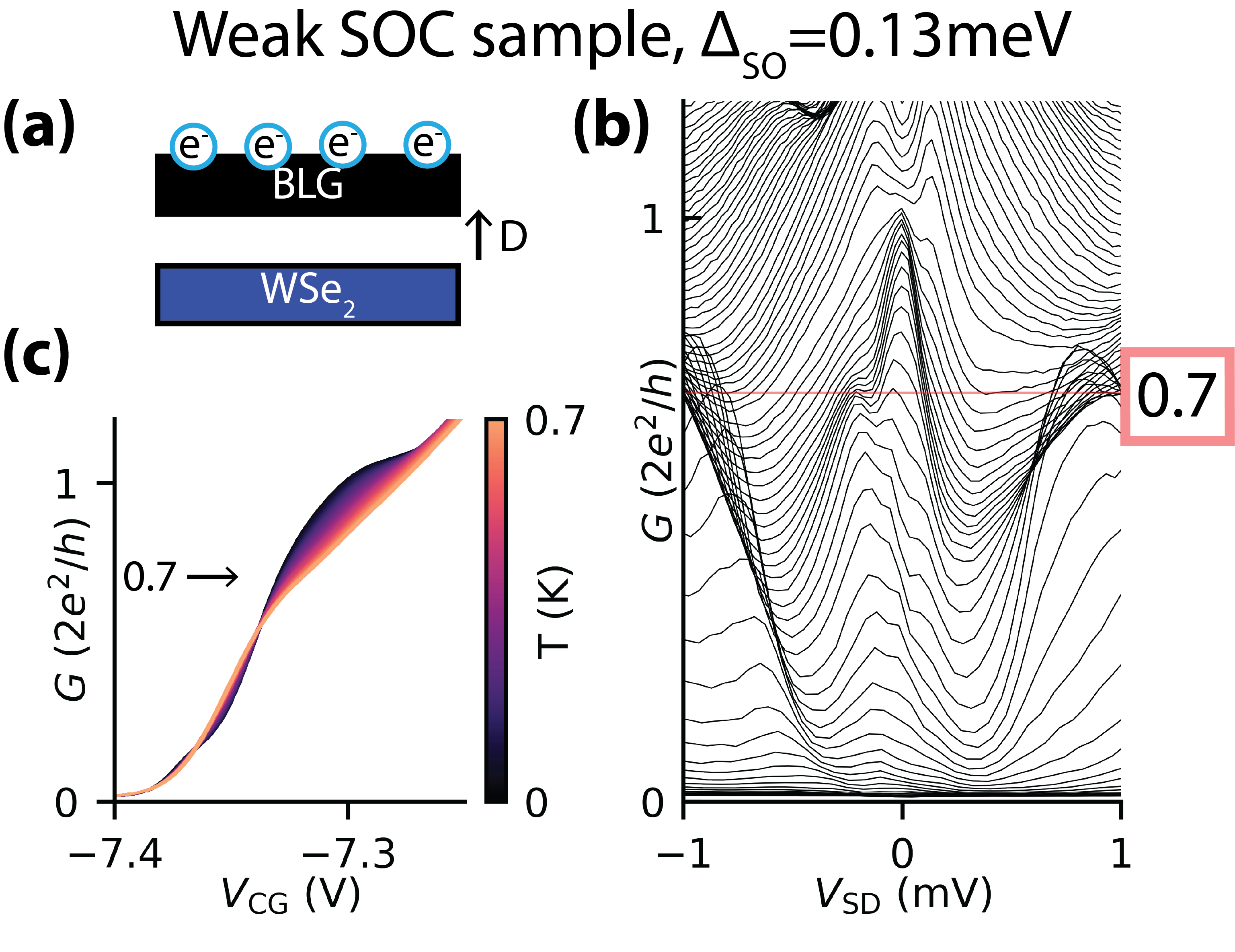}
	\caption{\textbf{(a)} Schematic stacking order of the weak SOC sample with indicated displacement field direction polarizing the electrons in the top graphene layer far from the TMD. \textbf{(b)} Differential conductance versus source-drain bias $V_\mathrm{SD}$ for varying $V_\mathrm{CG}$ showing indications of a  zero-bias anomaly and finite-bias plateaus at $0.7G_0$.   \textbf{(c)} Temperature dependence of conductance traces showing $G$ suppression at elevated temperatures.  }
    	\label{FigLowSOC}
\end{figure}

We observe indications of a 0.7 anomaly in a second BLG/WSe\textsubscript{2} sample with weak SOC ($\Delta_\mathrm{SO}$=\SI{0.13}{meV}), which is comparable to pristine BLG. When reversing the stacking order between WSe\textsubscript{2} and BLG while maintaining the displacement field direction, electrons are polarized in the layer remote from the WSe\textsubscript{2} (Fig.~\ref{FigLowSOC}a), resulting in weak SOC. The SOC is more than one order of magnitude lower compared to the strong SOC sample, shown in the main text. The weak SOC results in a small energy separation between the ground and first excited Kramers pair. As a result, there is no pronounced plateau at $G=2e^2/h$ (Fig.~\ref{FigLowSOC}b).

Despite this, we observe signatures consistent with the 0.7 anomaly: A characteristic temperature suppression (Fig.~\ref{FigLowSOC}b), a zero-bias anomaly, and a 0.7 shoulder at finite bias (Fig.~\ref{FigLowSOC}c). However, these features are significantly less distinct than in our high-SOC samples, in which states can be resolved individually and there is clear $G=2e^2/h$ conductance quantization. 
While this supports the universality of the spin--valley mechanism in BLG, the reduced quality could alternatively indicate more complex physics: when the energetic gap between states is small, all four states may contribute to the anomaly, as discussed for pristine BLG QPCs in Ref.~\cite{Gall2022}. The energetically closely spaced Kramers pairs in weak SOC samples prevent both clear conductance quantization at $2e^2/h$ and a definitive distinction between these scenarios. Strong spin--orbit coupling and the resulting state separation are hence essential for observing high-quality 0.7 anomalies and enabling unambiguous characterization of its properties.\\

\section{Device fabrication}

The heterostructures were fabricated by polymer-based dry transfer of mechanically exfoliated 2D materials. The samples in the main text (WSe\textsubscript{2}/BLG, strong SOC) and in the Supplementary Material  \cite{supplemental2025} (BLG/WSe\textsubscript{2}, weak SOC) consist of flakes with the following layer thicknesses:  

\begin{table}[h]
    \centering
    \renewcommand{\arraystretch}{1.2}
    \begin{tabular}{c|c|c}
        \hline
        \textbf{}  & \textbf{WSe\textsubscript{2}/BLG (main text)} & \textbf{BLG/WSe\textsubscript{2} (Fig.~\ref{FigLowSOC})} \\     \hline
        SOC  & strong ($\Delta_\mathrm{SO}=\SI{1.4}{meV}$) & weak ($\Delta_\mathrm{SO}=\SI{0.13}{meV}$) \\
        $D$&\SI{0.6}{V/nm}&\SI{0.7}{V/nm}\\
        
        \hline
        1 & hBN: \SI{21}{nm} & hBN: \SI{23}{nm} \\
        2 & WSe\textsubscript{2}: 5 layers & BLG \\
        3 & BLG  & WSe\textsubscript{2}: 3 layers \\
        4 & hBN: \SI{85}{nm} & hBN: \SI{42}{nm} \\
        5 & Graphite: \SI{3}{nm} & Graphite: \SI{25}{nm} \\
        \hline
    \end{tabular}
    \caption{Layer composition and thicknesses from top to bottom for both samples.}
    \label{tab:stack_thickness}
\end{table}

As both samples are identical to those in \cite{Gerber2025}, all fabrication-relevant information is included in that reference.

The displacement field values are estimated by $D=1/2\,(C_\mathrm{B}V_\mathrm{BG} - C_\mathrm{T}V_\mathrm{SG})$ using $\epsilon_r=3.24$ for hBN and $\epsilon_r=6$   for WSe\textsubscript{2} analog to Ref.  \cite{Gerber2025}.

\section{Measurement setup}
Measurements were performed in a He3/He4 dilution refrigerator with a base temperature of \SI{10}{mK}. The differential conductance was measured using a four-terminal configuration with an AC voltage excitation. The resulting current was converted to voltage using an IV amplifier and detected with a lock-in.

Despite four-terminal measurements, finite resistance contributions from the bulk BLG between the contacts remain.
We corrected the measured conductance $G_\mathrm{meas}$ by a constant series resistance of $G_\mathrm{S}= 1/\SI{0.8}{k\Omega}$ using the following equation:

 \begin{equation}
G=\frac{G_\mathrm{meas}G_\mathrm{S}}{G_\mathrm{S}-G_\mathrm{meas}}.
\end{equation} \label{eq:GQPC}

We determined this series resistance value by measuring the four-terminal resistance at the same back-gate voltage with both split gates and channel gates grounded. Under these conditions, the measured resistance is dominated by the bulk graphene resistance between contacts, and equals the series resistance $G_\mathrm{S}$.

\section{Extracting QPC properties}
\begin{figure}
	\includegraphics{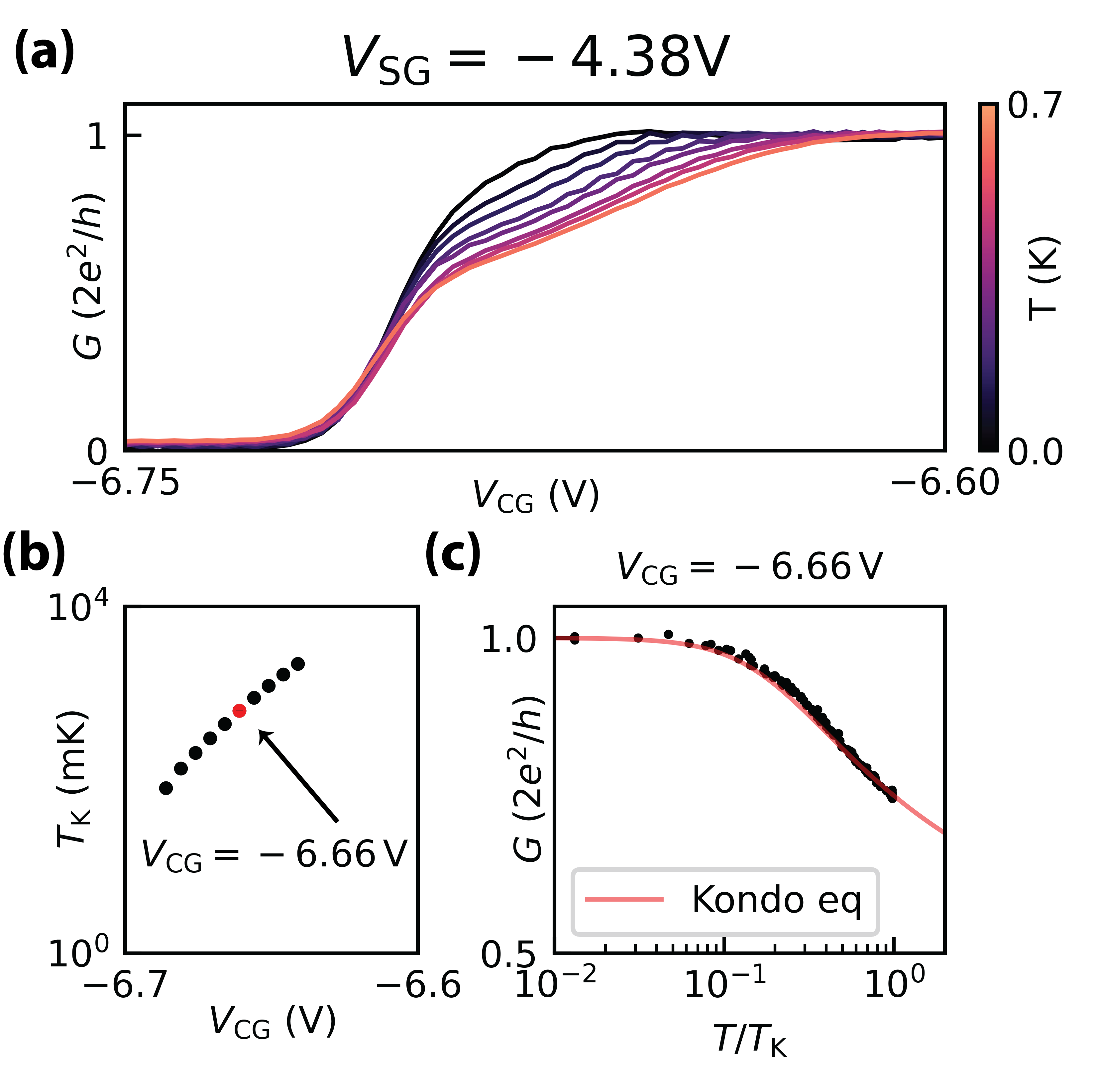}
	\caption{\textbf{(a)} Exemplary temperature-dependent conductance $G$ versus $V_\mathrm{CG}$. \textbf{(b)} Respective Kondo temperatures. \textbf{(c)} Conductance at $V_\mathrm{CG}=\SI{-6.66}{V}$ as a function of the temperature set in relation to the fitted Kondo temperature. The red curve shows the theoretical Kondo expression $G=G_\mathrm{0}\left[0.5\left(1+(2^{1/s}-1)(T/T_\mathrm{K})^2\right)^{-s}+0.5\right]$.  }
    	\label{FigKondoFit}
\end{figure}

Figure~\ref{FigKondoFit} illustrates the procedure for extracting Kondo temperatures (Fig.~\ref{FigKondoFit}b) from temperature-dependent conductance traces (Fig.~\ref{FigKondoFit}a). At each $V_\mathrm{CG}$, we extract linecuts of $G(T)$ and fit them to the modified Kondo expression $G=G_\mathrm{0}\left[0.5\left(1+(2^{1/s}-1)(T/T_\mathrm{K})^2\right)^{-s}+0.5\right]$ with $G_\mathrm{0}=2e^2/h$ and $s=0.22$, from which we extract the respective Kondo temperature $T_\mathrm{K}$ (Fig.~\ref{FigKondoFit}c). \\

To obtain an energy scale from the gate voltages, we perform finite-bias measurements in which the level splitting can be directly read off. The y-axis corresponds to the voltage drop across the QPC, which is the measured four-terminally and corrected  by the series resistance ($V_\mathrm{DC}=V_\mathrm{4term}-IG_\mathrm{S}$).

\begin{figure}
	\includegraphics{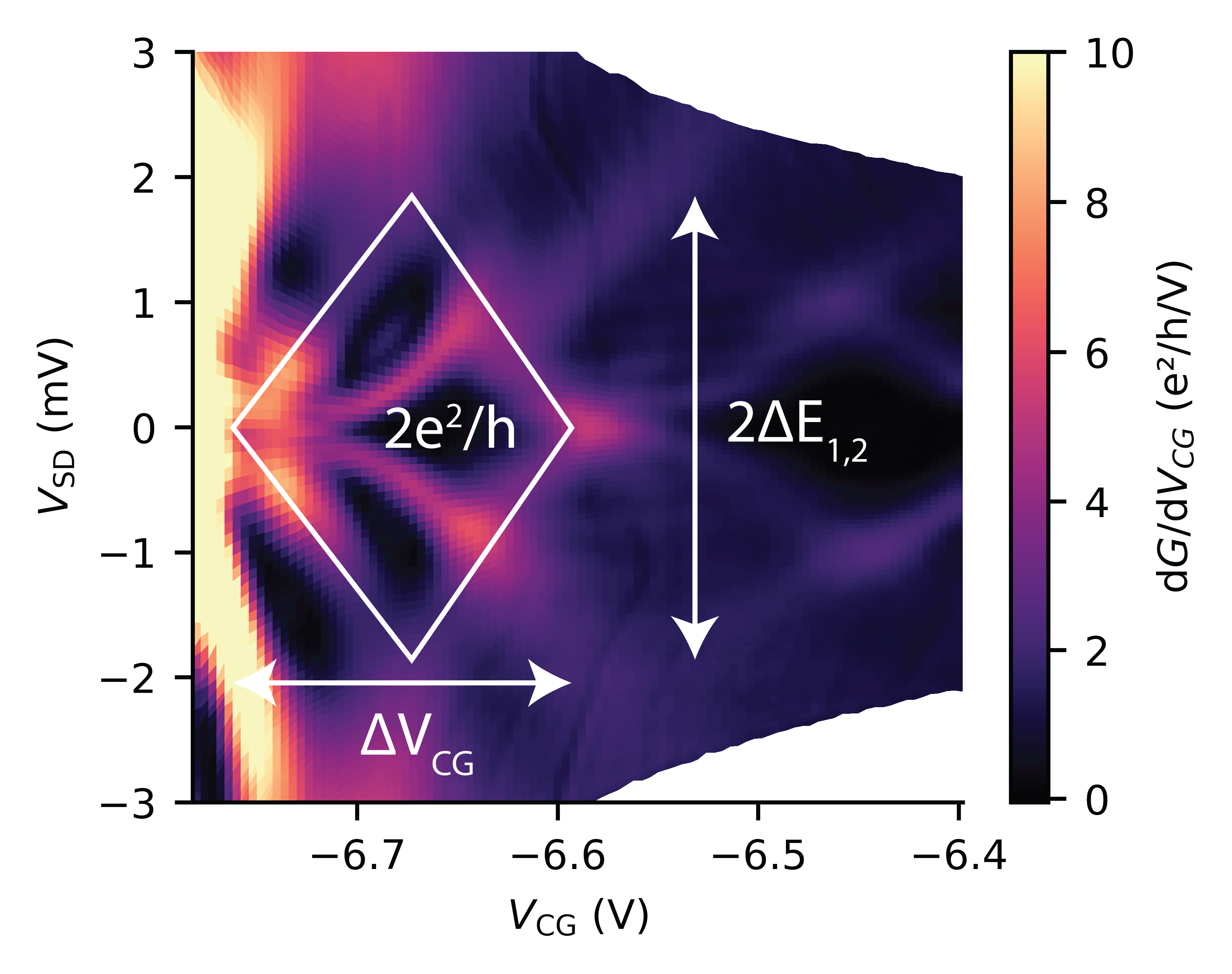}
	\caption{Finite-bias measurement and analysis of the first QPC subband of the sample shown in the main text. }
    	\label{FigDiamond}
\end{figure}

Due to the 0.7 anomaly, multiple diamonds appear around the first plateau (Fig.~\ref{FigDiamond}). We chose the outer diamond as an indicator of the subband spacing $\Delta E_\mathrm{1,2}$, analogously to the approach shown in \cite{Micolich2011}.\\

We address an apparent discrepancy in the spin--orbit gap value. From the level spectrum measurements on the present QPC, the subband spacing $\Delta E_{1,2}$ exceeds \SI{2.3}{meV}, which would suggest that $\Delta_\mathrm{SO}$ is larger than this value. However, we report $\Delta_\mathrm{SO}=\SI{1.4}{meV}$ in the main text, a value extracted from a different QPC on the same BLG/WSe\textsubscript{2} heterostructure \cite{Gerber2025}.

The two QPCs are fabricated on the identical heterostructure with the same twist angle and lattice orientation, but differ significantly in their interaction strength. The present QPC (\SI{50}{nm} effective channel width) exhibits strong electron interactions, evidenced by the prominent 0.7 anomaly. In contrast, the narrower QPC (\SI{30}{nm} effective width) from \cite{Gerber2025} shows much weaker interactions and no 0.7 anomaly. 

We consider $\Delta_\mathrm{SO}=\SI{1.4}{meV}$ from the narrow QPC more reliable because interaction effects can artificially enhance the apparent spin--orbit splitting \cite{Rozhansky2024}. Wider channels are known to enhance electron–electron interactions \cite{Hew2009, Kumar2012}, which not only drive the 0.7 anomaly but also cause an apparent increase in the measured spin–orbit gap through interaction effects. Therefore, we report the value from the weakly-interacting device as more representative of the intrinsic material property.\\


To extract the quantity $\hbar \omega_x$, we fitted the conductance traces with the transmission probability $T_n=(1+\exp[-2\pi(E-E_n)/\hbar\omega_{x,n}])^{-1}$. We converted the gate voltage axis $V_\mathrm{CG}$ into an energy axis using a constant factor of $\Delta E_{1,2}/\Delta V _\mathrm{CG}$. 

Some traces show a strong deviation from perfect transmission due to 0.7 anomaly shoulder formation. To extract the correct $\hbar \omega_x$, we therefore fit the conductance trace only below $G=0.4e^2/h$, as this region is unaffected by 0.7 anomaly effects.



\end{document}